\documentclass[preprint,12pt,authoryear]{elsarticle}

\usepackage{algorithm}
\usepackage{mdframed}
\usepackage{float}
\usepackage{tabularx}
\usepackage{xcolor}
\usepackage{amsmath}
\usepackage{changepage}
\usepackage{subcaption}
\usepackage{tcolorbox}
\usepackage[noend]{algpseudocode}
\usepackage{float}
\usepackage{amssymb}
\usepackage{multirow}
\usepackage{xcolor}
\makeatletter
\def\mathcolor#1#{\@mathcolor{#1}}
\def\@mathcolor#1#2#3{%
	\protect\leavevmode
	\begingroup
	\color#1{#2}#3%
	\endgroup
}
\makeatother

\usepackage{amssymb}

\journal{Journal of Applied Statistics}

\usepackage{xcolor}
\usepackage{hyperref}
\hypersetup{
	colorlinks= true,
	citecolor=blue,
	linkcolor=blue  
}

\usepackage{comment}
\usepackage{xstring}
\makeatletter
\AtBeginDocument{
	\let\oldref\ref
	\renewcommand{\ref}[1]{
		\IfBeginWith{#1}{fig:}%
		{{\color{blue}Figure~\oldref{#1}}}%
		\IfBeginWith{#1}{eqn:}%
		{{\color{blue}\oldref{#1}}}%
		{\IfBeginWith{#1}{tab:}{{\color{blue}Table~\oldref{#1}}}{}}}%
}
\makeatother

\begin{document}
	\begin{frontmatter}
		\title{Hierarchical Marketing Mix Models with Sign Constraints}
		
		\author{Hao Chen\fnref{label5}}
		\fntext[label5]{Corresponding Author}
		\ead{hao.chen@stat.ubc.ca}
		\author{Minguang Zhang}
		\ead{minguang.zhang@nielsen.com}
		\author{Lanshan Han}
		\ead{lanshan.han@nielsen.com}
		\author{Alvin Lim}
		\ead{alim2@precima.com}
		\address{Research \& Development, Precima, Chicago, IL 60606}

\begin{abstract} \label{sec:abstract}
Marketing mix models (MMMs) are statistical models for measuring the effectiveness of various marketing activities such as promotion, media advertisement, etc. In this research, we propose a comprehensive marketing mix model that captures the hierarchical structure and the carryover, shape and scale effects of certain marketing activities, as well as sign restrictions on certain coefficients that are consistent with common business sense. In contrast to commonly adopted approaches in practice, which estimate parameters in a multi-stage process, the proposed approach estimates all the unknown parameters/coefficients simultaneously using a constrained maximum likelihood approach and solved with the Hamiltonian Monte Carlo algorithm. We present results on real datasets to illustrate the use of the proposed solution algorithm.
\end{abstract}
		
\begin{keyword}
			Marketing Mix Model, Hierarchical Models, Constrained Regression Analysis, Hamiltonian Monte Carlo
\end{keyword}
\end{frontmatter}


\newpage
\section{Introduction}
Marketing activities, such as TV advertisement, discounting, direct mail, etc., are prevailing approaches for consumer packaged goods manufactures and service providers to enhance their brand awareness and product/service messaging to consumers in order to increase sales. It is therefore of tremendous interest to measure the return of investment (ROI) of those marketing activities. However, this is by no means an easy task, especially since it is very difficult, if not impossible at all, to conduct a controlled experiment. In fact, in practice, we usually collect sales, marketing, as well as other related data, often at weekly level, and then conduct statistical analysis to relate sales quantity to various marketing activities as well as other non-marketing factors. The statistical models constructed for this purpose are known as marketing mix models (MMMs). \\

There are often many complications in building a MMM. First, besides being affected by marketing activities, sales volume is also affected by many non-marketing factors, such as prices, holidays, seasonality, etc. These factors, while are not of interest themselves for the purpose of understanding effectiveness of marketing activities, need to be taken into consideration to properly measure the effects of marketing activities. Secondly, different marketing activities induce very different responses, which is technically more challenging. For instance, some marketing activities, such as promotional discounting, typically prompt an instant consumer response that vanishes as the activities end. Other marketing activities, such as TV advertising, may not elicit an immediate consumers response, but the carryover effect of the marketing activities might last beyond the active marketing period. These differences need to be captured in any applicable statistical models. Thirdly, the responses to any marketing activities are intrinsically heterogeneous along dimensions such as geography and product. For instance, the effect of a national TV advertisement may vary from one region to another due to geographical and demographic differences. It is essential to capture such heterogeneity in any applicable MMMs. Fourth, there often exists some sort of prior belief regarding the coefficients to be estimated. For example, while some marketing activities may not be effective, seldom do they have a negative impact on sales.  Mathematically, these kinds of belief are typically translated to linear inequality constraints on the coefficients, with sign constraints being probably the most common ones. In this paper, we present a comprehensive MMM that incorporates all the aforementioned considerations. \\

With all the complications discussed above, the resulting MMM often features nonlinear transformations with unknown parameters as well as inequality constraints on the parameters. Such a statistical model is certainly challenging to estimate. In practice, the estimation is often accomplished in multiple steps. For example, the practitioners often first estimate the parameters involved in the nonlinear transformations and then estimate the coefficients, followed by an adjustment process to ensure that the coefficients satisfying the required constraints. These multi-step process is not only complicated to implement and automate but could also lead to inaccurate estimation of the coefficients resulting in incorrectly measuring the effects of marketing activities on sales. Therefore, in this paper, we present a more systematized approach that allows us to estimate all the unknown parameters simultaneously, while ensuring that all the constraints are satisfied. \\

The rest of this paper is organized as follows. In Section~\oldref{sec:MMM}, we lay out a detailed discussion on the features of marketing mix models and then provide a literature review. In Section~\oldref{sec:spec}, we present model specifications, including details on how to capture carryover, shape and scale effects. In Section~\oldref{sec:hmc}, we present our estimation approaches. Results from some numerical studies and analyses on real datasets are reported in Section~\oldref{sec:exp} and Section~\oldref{sec:real}, respectively, followed by concluding remarks in Section~\oldref{sec:con}.

\section{Marketing Mix Models}\label{sec:MMM}

As was mentioned earlier, different marketing activities often generate different responses from consumers. Among all the marketing activities, advertisements are the ones that introduce unique challenges. The reasons are twofold. First, advertisements typically generate long lasting but decaying effects that go beyond the time period of active advertisement. Therefore, when we study the response from the advertisements from week to week, it looks as if a portion of the investment from previous weeks still generate response in the current week. This carryover phenomenon is known as ``adstocking'' in marketing practice \citep{bickart1993carryover}. More specifically, we typically use targeted rating points (TRPs) \citep{surmanek1996media} to measure the level of activity for advertisements. We study a period of $w$ consecutive weeks, labeled by $t=1,\cdots,w$. We denote the TRP of an advertisement in week $t$ by $x_t$. Due to the carryover effect, the \emph{effective} TRP in week $t$ is given by:
 $$\widetilde{x}_t \, = \, c(x_1,\cdots,x_t; \theta),$$
 where $\theta$ is an unknown parameter. Note that the carryover effect from weeks earlier than the study period can be considered similarly, but we ignore such effects for simplicity of demonstrating our approach. In this paper, we consider a specific format of $c(x_1,\cdots,x_t; \theta)$ given by
 \begin{equation} \label{eqn:carryover}
 c(x_1,\cdots,x_t; \theta) \, = \, \sum_{\tau=0}^{\ell-1} \alpha^{\tau} x_{t-\tau}, \,\, \forall \, t=\ell,\cdots,w;
 \end{equation}
with $\alpha \in (0,1)$, referred to as the decay rate hereafter. This is to say that the carryover effect becomes negligible after $\ell$ weeks, decays by an unknown constant factor $\alpha$ each week, and is additive. In practice, one either determines the decay rate using rule-of-thumb based on experience, or estimate $\alpha$ in a pre-processing step before the effectiveness of the marketing activities are estimated. Ideally, we should let the data speak for itself and estimate the decay rates together with marketing effectiveness simultaneously. \\

\begin{figure}[H]
	\centering
	\includegraphics[width=0.91\textwidth, scale=1.0]{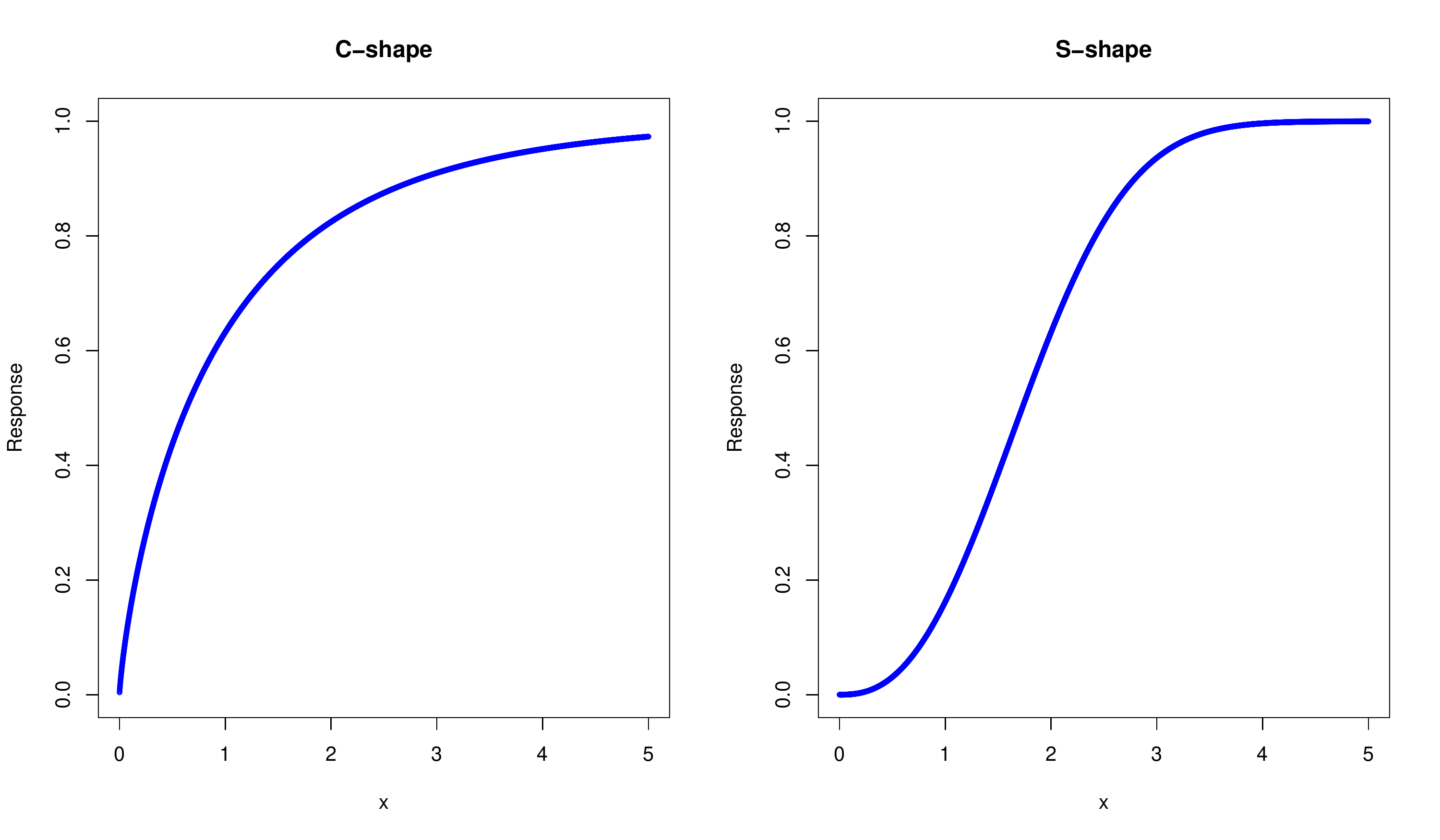}
	\caption{Illustrating plots of C-shape and S-shape} \label{fig:cs_shape}
\end{figure}

Another level of complexity regarding advertisements is that the effects are in general nonlinear. Specifically, it is widely recognized that all advertisements are subject to a so-called ``saturation" phenomenon. Generally speaking, saturation refers to the fact that while the response still increases as the TRP increases, the rate slows down as the advertisement TRPs continue to increase. This is because of the fact that the targeted population exposed to the advertisement is finite. We use response functions to mathematically link the effectiveness of advertisements and TRPs. Due to the saturation phenomenon, a response function is typically either a C-shape (concave increasing) or S-shape (non-concave increasing) as illustrated in Figure~\oldref{fig:cs_shape}.

We propose to use the cumulative distribution function (CDF) of Weibull distribution to capture these two different shapes. The Weibull CDF, taking two parameters, is given as follows,
\begin{equation} \label{eqn:shape}
s(\widetilde x; \lambda, k) \, = \, 1 - \exp\left(- \left(\frac{\widetilde x}{\lambda}\right)^k\right).
\end{equation}
The response function of the advertisement is therefore
\begin{equation} \label{eqn:r}
r(\widetilde x; \beta,\lambda, k) \, = \, \beta s(\widetilde x; \lambda, k),
\end{equation}
with $\beta$ being the unknown coefficient of effectiveness, and $\lambda$, $k$ being unknown nonlinear transformation parameters. We refer to $\lambda$ and $k$ as the shape and scale parameters, respectively, hereafter. Note that all $ \alpha, k, \lambda  $ require estimation in practice. \\


The third layer of complications comes from the common belief that the advertisements, while may be completely ineffective, at least will not affect sales negatively. Mathematically, this can be translated to an inequality constraint, i.e., $\beta\geq 0$. Traditionally, statistical estimations are often unconstrained or under only equality constraints. The Inclusion of inequality constraints impose significant challenges, especially under a hierarchical structure, which we will elaborate in the next paragraphs. In fact, most commercially available statistical software packages do not allow us to explicitly impose inequality constraints on the parameters to be estimated. Therefore, practitioners often need to apply some heuristics to ``correct" the signs after the coefficients are estimated. In the propose approach, we explicitly impose these constraints, and therefore no ad-hoc ``corrections" are needed after estimation. \\

The fourth layer of complications lies in the fact that there is intrinsic heterogeneity along geography and product dimensions. For example, the response to an advertisement can vary from one geographical region to another, and hence so do the coefficients of effectiveness. In the meantime, we often believe that those coefficients, while different from each other, behave like having a common coefficient adjusted by a random coefficient following a Normal distribution with $0$ mean and unknown variance. This comes under the framework of mixed effect models, which will be discussed in the following sections. Mathematically, let $\nu$ denote different geographical regions indexed by $\nu=1,\cdots,g$. For each region $\nu$, the coefficient of effectiveness $\beta_\nu \overset{iid}{\thicksim} \mathcal N(\beta, \eta^2)$, with $\mathcal N (\beta,\eta^2)$ referring to a Normal distribution with mean $\beta$ and variance $\eta^2$. \\

With all the added tiers of complexity, the MMM is a highly challenging statistical model to estimate. In this research, we discuss learning the unknown parameters from both a frequentist perspective via maximum likelihood estimation (MLE), and Bayesian viewpoint using the Hamiltonian Monte Carlo (HMC) approach \citep{hmc}. HMC is a variant of the traditional Metropolis-Hastings algorithm \citep{chib1995understanding}, which belongs to the family of Markov chain Monte Carlo (MCMC) algorithms. The benefits of HMC over the Metropolis-Hastings algorithm will be discussed in Section \oldref{sec:hmc}.\\

The research on MMMs dates back to 1960's, with a conceptual framework being laid out in \cite{Borden64} when it was originally referred to as the 4Ps (Product, Price, Place, Promotion). Some early developments in the 1970's can be found in \cite{lambin1972computer} and \cite{little1975brandaid}. It became widely known after being included in a classical textbook \citep{McCarthy78}. Traditionally, the regression analysis is carried out using a frequentist paradigm via MLE. In recent years, marketing mix modeling has received renewed interest due to the emergence of advertising channels such as paid search, digital coupons, etc., as well as progresses made in statistical methods such as Bayesian inference using MCMC approach and computation facilities allowing large-scale parallelization such as the use of graphical processing units (GPUs). The research and challenges are summarized in a recent survey \citep{CPerry17}. This paper largely adopts the framework by \cite{google}, while expanding it to incorporate heterogeneity in marketing response \citep{SWJCKoehler17} as well as allowing sign constraints on the coefficients to be estimated.

\section{Hierarchical Marketing Mix Model} \label{sec:spec}
In this section, we provide detailed statistical models for marketing mix modeling. We will begin with the base model and then introduce hierarchical structure as well as constraints.
\subsection{Base Model}
In this section, we present the base marketing mix model without hierarchical structure. Without loss of generality, we assume there are $d$ independent variables in total, and the first $m$ variables, denoted as $ x_i, i=1,\cdots,m  $, have carryover, shape and scale effects. The remaining $n = d-m$ variables are nuisance variables (representing non-marketing factors), denoted as $z_j, j=1,\cdots, n$. The dependent variable is the sales quantity (possibly transformed) denoted as $ y $. Observations have been collected from $w$ consecutive weeks, ordered chronologically, and indexed by $t=1, \cdots, w$. The available dataset is depicted in Table~\oldref{tab:data}. \\

\begin{table}[http!]
	\centering
	\caption{Available data from $w$ weeks for the base model}\label{tab:data}
	\begin{tabular}{|c|cccc|cccc|}
		\hline
		$y_1$ & $x_{1, 1}$ & $ x_{1, 2} $ & $ \cdots $ & $ x_{1, m} $ & $z_{1, 1}$ & $ z_{1, 2} $ & $ \cdots $ & $ z_{1, n} $ \\
		$\vdots$ & $ \vdots $ & $  \vdots $ & $ \vdots $ & $ \vdots $ & $ \vdots $ & $  \vdots $ & $ \vdots $ & $ \vdots $ \\
		$y_w$ & $x_{w, 1}$ & $ x_{w, 2} $ & $ \cdots $ & $ x_{w, m} $ & $z_{w, 1}$ & $ z_{w, 2} $ & $ \cdots $ & $ z_{w, n} $ \\
		\hline
	\end{tabular}	
\end{table}

In the dataset, $x_{t,i}$ is the organic value of independent variable $x_i$ in week $t$ without considering the carryover, shape and scale effects at week $ t $. As we have discussed in the previous section, we let $\alpha_i$, $k_i$, and $\lambda_i$ be the decay, shape, and scale parameters, respectively, of variable $x_i$. For simplicity, we assume the maximum carryover period $\ell$ is known and the same for all different marketing campaigns. In practice, $\ell$ can be chosen to be large enough so that carryover effects beyond $\ell$ weeks are negligible. After taking into account the net effect of carryover, shape and scale transformations on each independent variable $x_i$, $i=1,\cdots,m$ in week $t=\ell, \cdots, w$, the response function becomes:
\begin{equation} \label{eqn:r_css}
r(x_{t-\ell+1,i},\cdots,x_{t,i}; \beta_i,\lambda_i, k_i, \alpha_i) \, = \, \beta_i s_i(c_i(x_{t-\ell+1,i},\cdots,x_{t,i};\alpha_i);k_i,\lambda_i),
\end{equation}
where $ c(\cdot) $ and $ s(\cdot) $ are defined in (\oldref{eqn:carryover}) and (\oldref{eqn:shape}), respectively. Therefore, the overall base model is given by:
\begin{equation} \label{eqn:base_model}
y_t \, =\, \sum_{i=1}^m  \beta_i s_i( c_i(x_{t-\ell+1,i},\cdots, x_{t,i} ;\alpha_i);k_i, \lambda_i) \, + \, \sum_{j=1}^n \gamma_j z_{t,j}+ \epsilon_t, \,\,
\end{equation}
for all $\, t=\ell,\ell+1,\cdots, w,$ where $\epsilon_t \sim N(0,\sigma^2)$ and is independent for all $t=\ell,\ell+1,\cdots,w$. The unknown parameters $\alpha_i$'s, $k_i$'s, $\lambda_i$'s, $\beta_i$'s, $\gamma_j$'s as well as $\sigma^2$ require estimation in practice. Note that one of the $\gamma_j$'s can be an intercept.

\subsection{Extension to the Hierarchical Model}
The model in (\oldref{eqn:base_model}) represents a linear model after the carryover, shape and scale effects are considered. To account for heterogeneity along different geographical dimensions, it is often necessary to incorporate hierarchical structures, which leads to general linear hierarchical models. We are particularly interested in hierarchical models with mixed effects, in which some or all of the independent variables have a hierarchy to account for the heterogeneity across sub-populations such as different regions using random coefficients. Moreover, we also propose to have sign constraints on some of the coefficients to be consistent with our business knowledge and common sense. For example, the coefficients for marketing activities should, in general, be non-negative. Compared to Table~\oldref{tab:data}, we further assume that the data contain an additional layer of regions, indexed by $ \nu = 1, 2, \cdots, g$. The available data for hierarchical modeling is given in Table~\oldref{tab:data2}. \\

\begin{table}[http!]
	\centering
	\caption{Available data from $ g $ regions. Each has $w$ weeks marketing data.}\label{tab:data2}
	\begin{tabular}{|c|c|cccc|cccc|}
		\hline
$\nu = 1$	 &	 $y_{1,1}$ & $x_{1, 1,1}$ & $ x_{1, 2, 1} $ & $ \cdots $ & $ x_{1, m, 1} $ & $z_{1, 1, 1}$ & $ z_{1, 2,1} $ & $ \cdots $ & $ z_{1, n, 1} $ \\
$\vdots$	 &	$\vdots$ & $ \vdots $ & $  \vdots $ & $ \vdots $ & $ \vdots $ & $ \vdots $ & $  \vdots $ & $ \vdots $ & $ \vdots $ \\
$\nu = 1$	 &		$y_{w, 1}$ & $x_{w, 1, 1}$ & $ x_{w, 2, 1} $ & $ \cdots $ & $ x_{w, m, 1} $ & $z_{w, 1, 1}$ & $ z_{w, 2, 1} $ & $ \cdots $ & $ z_{w, n, 1} $ \\
		\hline
$\vdots$	 &	 $\vdots$ & $\vdots$ & $ \vdots $ & $ \ldots $ & $ \vdots $ & $\vdots$ & $ \vdots $ & $ \cdots $ & $ \vdots $ \\
$\vdots$	 &		$\vdots$ & $ \vdots $ & $  \vdots $ & $ \vdots $ & $ \vdots $ & $ \vdots $ & $  \vdots $ & $ \vdots $ & $ \vdots $ \\
$\vdots$	 &		$\vdots$ & $\vdots$ & $ \vdots $ & $ \ldots $ & $ \vdots $ & $\vdots$ & $ \vdots $ & $ \ldots $ & $ \vdots $ \\
		\hline
$\nu = g$	 &	  $y_{1, g}$ & $x_{1, 1, g}$ & $ x_{1, 2, g} $ & $ \cdots $ & $ x_{1, m, g} $ & $z_{1, 1, g}$ & $ z_{1, 2, g} $ & $ \cdots $ & $ z_{1, n, g} $ \\
$\vdots$	 &		$\vdots$ & $ \vdots $ & $  \vdots $ & $ \vdots $ & $ \vdots $ & $ \vdots $ & $  \vdots $ & $ \vdots $ & $ \vdots $ \\
$\nu = g$	 &		$y_{w, g}$ & $x_{w, 1, g}$ & $ x_{w, 2, g} $ & $ \cdots $ & $ x_{w, m, g} $ & $z_{w, 1, g}$ & $ z_{w, 2, g} $ & $ \cdots $ & $ z_{w, n, g} $ \\
\hline
	\end{tabular}	
\end{table}

We let $\mathcal H_{\beta}$ and $\mathcal H_{\gamma}$ be the set of indices of $\beta$-variables with sign constraints and $\gamma$-variables with sign constraints, respectively. We define $\overline {\mathcal H_{\beta}}$ and $\overline {\mathcal H_{\gamma}}$ be the complement of $\mathcal H_{\beta}$ and $\mathcal H_{\gamma}$ in $\{1,\cdots,m\}$ and $\{1,\cdots, n\}$, respectively. Without loss of generality, we assume all the sign constrains are nonnegative constraints. The hierarchical model is therefore:

\begin{equation}\label{eqn:full_model}
\begin{array}{rcl}
y_{t,\nu} & = & \displaystyle {\sum_{i\in \mathcal H_\beta}  \beta_{i,\nu} s_i( c_i(x_{t-\ell+1,i,\nu},\cdots, x_{t,i,\nu} ;\alpha_i);k_i, \lambda_i)}\\[5pt]
   && \displaystyle{ \,+ \,\sum_{i\in \overline{ \mathcal H_\beta} } \beta_{i, \nu} s_i( c_i(x_{t-\ell+1,i,\nu},\cdots, x_{t,i,\nu} ;\alpha_i);k_i, \lambda_i)  } \\[5pt]
   && \displaystyle{ \, + \, \sum_{j\in \mathcal H_{\gamma}} \gamma_{j,\nu} z_{t,j,\nu}\, + \, \sum_{j\in \overline{\mathcal H_{\gamma} } } \gamma_{j, \nu} z_{t,j,\nu} \, + \,\epsilon_{t,\nu} } , \, \forall \, t=\ell,\cdots, w; \nu = 1, \cdots,  g\\[10pt]
\beta_{i,\nu} & \overset{iid}{\thicksim} &  N(\beta_j, \eta_i^2), \,\forall\, i =1, \cdots, m \\[5pt]
\gamma_{j,\nu} & \overset{iid}{\thicksim} &  N(\gamma_j, \xi_j^2), \,\forall\, j =1, \cdots, n \\[5pt]
\epsilon_{t,\nu} & \overset{iid}{\thicksim} &  N(0, \sigma^2)\\[5pt]

\beta_{i,\nu} & \geq & 0, \, \forall \, i \in \mathcal H_{\beta}  \\[5pt]
\beta_{i} & \geq & 0, \, \forall \, i \in \mathcal H_{\beta} \\[5pt]

\gamma_{j,\nu} & \geq & 0, \, \forall \, j \in \mathcal H_{\gamma} \\[5pt]
\gamma_{j} & \geq & 0, \, \forall \, j \in \mathcal H_{\gamma} \\[5pt]
\end{array}
\end{equation}
In this model, we assume the carryover, shape and scale parameters are the same across different regions, while they can vary across different marketing activities. In theory, we could also allow them to vary across sub-populations. However, this may lead to significantly enlarged parameter space and lead to identifiability issues.  \\

Given the hierarchical model in (\oldref{eqn:full_model}), it is obvious that we have the following parameters that need to be estimated:
\begin{itemize}
	\item Carryover parameters: $ \alpha_1, \cdots, \alpha_m $
	\item Shape parameters: $ k_1, \cdots, k_m $
	\item Scale parameters: $ \lambda_1, \cdots, \lambda_m $
	\item The means of fixed regression parameters: $ \beta_1, \cdots, \beta_m; \gamma_1, \cdots, \gamma_n $
	\item The variances of fixed regression parameters: $ \eta_1^2, \cdots, \eta_m^2; \xi_1^2, \cdots, \xi_n^2 $
	\item The random regression parameters: $ \beta_{i, \nu}, \gamma_{j, \nu}$, where $ i=1,\cdots, m $, $ j=1, \ldots, n $ and $ \nu = 1, \cdots, g $.
	\item The variance of the model, $ \sigma^2 $
\end{itemize}
In addition, the parameters are constrained such that $ 0 \le \alpha_i <1 $, $ k_i >0, \lambda_i >0$, $  \beta_i \geq 0 $, $ \gamma_j \geq 0 $, $ \beta_{i, \nu} \geq 0$, and $  \gamma_{j, \nu} \geq 0,$ for all $i \in \mathcal{H}_\beta, j \in \mathcal H_\gamma$. With the sign constraints, the model estimation becomes more challenging no matter which estimation approach we take. When maximum likelihood estimation is applied, the maximization problem is an inequality constrained nonlinear nonconvex optimization problem. When we adopt a Bayesian inference paradigm, the major challenge is to manage the computation efficiency as well as handling the constraints. We present details regarding parameter estimation of the proposed MMM in the next section.

\section{Parameter Estimation of MMMs}\label{sec:hmc}
In general, there are two different parameter estimation methods: the first one is the frequentist paradigm via maximum likelihood (ML) estimation \cite{larson1969introduction}. The basic idea is to take the log likelihood as a function of the unknown parameters, and then find estimates such that they maximize the log likelihood function. Since the parameters are constrained, it will be further viewed as a non-linear constrained optimization problem. The second method is the Bayesian paradigm via Bayes' theorem. For most practical problems, direct sampling from the posterior distribution of the unknown parameters is unavailable. Therefore, the parameters are inferred using Markov chain Monte Carlo (MCMC) \citep{gilks1995markov}, in which one builds a Markov chain whose stationary distribution is the posterior distribution. Then one collects samples after burn-in. More details about these general ideas can be found in nearly any modern statistical inference textbook, for example \cite{larson1969introduction} and \cite{rao1973linear}. \\

The frequentist paradigm is straightforward, but our parameter space is high dimensional and imposes much difficulty for constrained optimization methods to produce reasonable estimates. This is confirmed in the simulation study in Section \oldref{sec:exp}. Therefore, we propose to use the Hamiltonian Monte Carlo (HMC) algorithm \citep{hmc} to infer the unknown parameters. HMC is a variant of the Metropolis-Hastings method \citep{chib1995understanding}, which is one of the most popular MCMC methods. HMC follows the framework of Metropolis-Hastings method, but HMC proposes candidates following the Hamiltonian dynamics. In the rest of this section, we will discuss both ML and HMC approaches.\\
\subsection{Maximum Likelihood Estimation of MMMs}
To facilitate the MLE approach, we first examine the likelihood function. We notice that, due to the existence of sign constraints on $\beta_{i, \nu}$ for all $i\in \mathcal H_\beta$, the probability density function (PDF) of $\beta_{i,\nu}$ given $\beta_i, \eta_i^2$ should be considered as a one-sided truncated Normal distribution. That is, for all $i \in \mathcal H_\beta$ it is given by
\begin{equation*}
f(\beta_{i,\nu}|\beta_i, \eta_i^2) = \zeta_{\beta,i}(\beta_i) \frac{1}{\sqrt{2\pi}\eta_i} \exp\left(-\frac{1}{2}\left(\frac{\beta_{i, \nu} - \beta_i}{\eta_i}\right)^2\right),
\end{equation*}
which is the PDF of a Normal distribution $N(\beta_i, \eta_i^2)$ with a $\beta_i$-\emph{dependent} scaling factor
$$\zeta_{\beta,i}(\beta_i) \, = \, \frac{1}{\eta_i (1- \Phi(-\frac{\beta_i}{\eta_i}))},$$
where $\Phi(\cdot)$ is the cumulative distribution function (CDF) of standard Normal distribution, i.e.,
$$\Phi(\omega) \, = \, \frac{1}{2}\left(1+ \mbox{erf}\left(\frac{\omega}{\sqrt 2}\right)\right).$$
The dependence of the scaling factor, in fact, has profound implications. In particular, it is not correct to ignore this scaling factor when maximizing the likelihood function, and hence the likelihood function is fundamentally different to the one for traditional unconstrained linear hierarchical models. Similarly, the PDF of $\gamma_{j,\nu}$ given $\gamma_j, \xi_j^2$, for all $j\in \mathcal H_\gamma$ is given by
\begin{equation*}
f(\gamma_{j,\nu}|\gamma_j, \xi_j^2) = \zeta_{\gamma,j}(\gamma_j) \frac{1}{\sqrt{2\pi}\xi_j} \exp\left(-\frac{1}{2}\left(\frac{\gamma_{j, \nu} - \gamma_j}{\xi_j}\right)^2\right),
\end{equation*}
with
$$\zeta_{\gamma,j}(\gamma_j) \, = \, \frac{1}{\xi_j (1- \Phi(-\frac{\gamma_j}{\xi_j}))}. $$
Let $\Theta$ denoted all the unknown parameters to be estimated. With $r_i$ analogously defined as in (\oldref{eqn:r_css}), we have
\begin{equation*}
\resizebox{\textwidth}{!}{$f(y_{t, \nu}|\Theta) = \frac{1}{\sqrt{2\pi}\sigma} \exp\left(-\frac{1}{2}\left(\frac{y_{t, \nu} -  (\sum_{i=1}^{m} r_i(x_{t, i, \nu}; \beta_{i, \nu}, \alpha_i, \lambda_i, k_i) + \sum_{j=1}^{n}\gamma_{j, \nu}z_{t, j, \nu}) }{\sigma}\right)^2\right)$},
\end{equation*}
for $t=\ell, \cdots, w$. Therefore, the joint likelihood function is given by:
\begin{equation} \label{eqn:likelihood}
L(\Theta) \, = \, \left(\prod_{t=\ell}^{w} \prod_{\nu=1}^{g} f(y_{t, \nu}|\Theta)\right) \times
\left(\prod_{i=1}^{m}\prod_{\nu=1}^{g}f(\beta_{i, \nu} | \beta_i, \eta_i^2 )\right)
\times
\left(\prod_{j=1}^{n}\prod_{\nu=1}^{g}f(\gamma_{j, \nu} | \gamma_j, \xi_j^2)\right).
\end{equation}
The ML approach hence leads to the following constrained optimization problem:
\begin{equation}\label{eqn:ML_constrained}
\begin{array}{rll}
\max_{\Theta} & \ln(L(\Theta)) \\ [5pt]
\mbox{s.t} & \beta_i \, \geq \, 0 & i \in \mathcal H_\beta \\
& \beta_{i, \nu} \, \geq \, 0 & i \in \mathcal H_\beta, \nu=1,\cdots,g \\
& \gamma_j \, \geq \, 0 & j \in \mathcal H_\gamma \\
& \gamma_{j,\nu} \, \geq \, 0 & j \in \mathcal H_\gamma , \nu = 1,\cdots, g.
\end{array}
\end{equation}
As we can see, the objective function in (\oldref{eqn:ML_constrained}) is highly nonlinear and non-convex. On the other hand, the constraints are relatively simple. We can apply different optimization algorithms to solve this problem, although it is typically impossible to find a global optimal solution of (\oldref{eqn:ML_constrained}). Generally speaking, optimization algorithms, almost all of which are of iterative nature, can be categorized into three categories. In the first category of the algorithms, only first order information (gradient) of the objective function and/or the constraints is utilized. These algorithms are typically cheap in terms of computational time and space complexity at each iteration, but often requires a lot of iterations for convergence. In the second category of the algorithms, second order information (Hessian matrix) of the objective function is utilized. These algorithms are more expensive at each iteration, especially when the number of variables is large, but often require less iterations to converge. The third category of algorithms is somewhere in between. They aim to approximate the Hessian matrix in a less expensive way compared to obtaining the exact Hessian matrix. These algorithms typically converge in a reasonable number of iterations. Due to the complexity of the objective in (\oldref{eqn:ML_constrained}) as well as the potential large number of variables, we opt for this third class of algorithms. In particular, we apply two different algorithms: one is the limited memory version of bounded Broyden-Fletcher-Goldfarb-Shanno (L-BFGS-B) \citep{zhu1997algorithm}, and the other is sequential quadratic programming (SQP) \citep{boggs1995sequential} algorithm. Numerical results will be provided in Section \oldref{sec:exp}.

\subsection{Hamiltonian Monte Carlo Approach}
Hamiltonian Monte Carlo is a type of MCMC approach which uses Hamiltonian dynamics to propose new random samples. Traditional Gaussian random walk Metropolis-Hastings algorithms typically use a one-dimensional Normal proposal. More specifically, let $\Theta = (\theta_1, \cdots, \theta_d)^T$ be the vector of all parameters. A random walk proposed for the $j$-th parameter is drawn from a Normal distribution $$ \theta^{*}_j \sim N(\theta_j, \sigma_j^2 | \theta_1, \cdots, \theta_{j-1}, \theta_{j+1}, \cdots, \theta_{d}),$$ where $ \sigma^2_j $ is tuned to ensure the acceptance rate is about 20 - 40\% \citep{rosenthal2011optimal}. As we can see, at each iteration, it needs to independently propose a candidate for each variable. The drawbacks are obvious: first of all, one needs independent proposals for each variable and then combine and evaluate $ \Theta^{*} $ collectively, usually leading to a high rejection rate and inefficiency for high-dimensional problems. Secondly, it is mathematically tedious as one needs to analytically derive the conditional posterior distribution for each unknown parameter. The remedy proposed is the HMC approach, which is able to propose multi-dimensional candidate at one shot. To do so, we follow the so-called Hamiltonian dynamics, which was first studied by physicists, and was later borrowed by statisticians. Hamiltonian dynamics \citep{dirac1950generalized} describes a frictionless puck that slides over a surface of varying height. The state of the system consists of the position (given by a vector $z$) of the puck and the momentum of the puck (given by a vector $v$). The potential energy, $U(z)$ is viewed as a function of $ z $ and the kinetic energy, $K(v) = |v|^2/(2s)$, where $ s $ is the mass of the puck. If the puck encounters a rising slope, the puck's momentum allows it to continue, with its kinetic energy decreasing and its potential energy increasing, until the kinetic energy is zero when it will slide back. Let the Hamiltonian be defined as $H(z, v) = U(z) + K(v)$. This dynamics is described by the following differential equations:
\begin{eqnarray*}
	\frac{\partial z_i}{\partial t_i} &=& \frac{\partial H}{\partial v_i}, \\
	\frac{\partial v_i}{\partial t_i} &=& -\frac{\partial H}{\partial z_i}. \\
\end{eqnarray*}

When applying the HMC algorithm, we let $z$ be the vector of unknowns, i.e., $z=\Theta$, and let $v$ be an auxiliary vector of the same dimension as $\Theta$. Let $K(v) = \frac{1}{2}v^Tv$. Let $U(\Theta) = -\ln (P(\Theta))$ with $P(\Theta)$ being the posterior PDF of the unknowns up to a multiplicative constant. Typically, $P(\Theta)$ is the product of prior distribution and likelihood function. We use a leapfrog procedure \citep{hmc}, which is an enhancement to the explicit Euler's method \citep{jain1979numerical}. At a given time $\tau$, the leapfrog method compute $\Theta(\tau+\Delta \tau)$ and $v(\tau + \Delta \tau)$ by
\begin{eqnarray}
	v(\tau+\Delta \tau/2) &=& v(\tau) - \frac{\Delta \tau}{2}\nabla P(\Theta)|_{\Theta=\Theta(\tau)} \nonumber \\[5pt]
	\Theta(\tau+\Delta \tau) &=& \Theta(\tau) + \Delta \tau v(\tau+\Delta \tau/2) \label{eqn:full_step_q} \\[5pt]
	v(\tau+\Delta \tau) &=& v(\tau + \Delta \tau/2)-\nabla P(\Theta)|_{\Theta=\Theta(\tau+\Delta \tau)} \nonumber
\end{eqnarray}
As we can see, we start from the current $v$ and $z$ and then first updating $v$ a half step, then the position $z$ a whole step, and then finish by updating $v$ the other half of the step. The magnitude of $ \Delta \tau$ is called the step size. Note that equations (\oldref{eqn:full_step_q}) can be repeated for $\kappa$ times, to obtain $\Theta(\tau+\kappa \Delta \tau)$ and $v(\tau + \kappa \Delta \tau)$. We then let $\Theta^* = \Theta(\tau+\kappa \Delta \tau)$ be the proposal. It is worth pointing out that $\Delta \tau$ and $\kappa$ are two important user-defined parameters that one needs to carefully tune them to make the overall acceptance rate close to HMC's optimal acceptance rate $0.65$ \citep{hmc}.

Let $ \Theta^{(\imath)} $ collectively denote the unknown parameters at iteration $ \imath $. $ \Theta^{*} = \Theta(\tau+ \kappa \Delta \tau)$ is the proposal generated by repeating the leapfrog process $\kappa$ times with $\Theta(\tau) = \Theta^{(\imath)}$ and $v(\tau)$ being a random sample from multivariate normal distribution $N(\mathbf 0_d, \mathbf I_d)$, where $\mathbf 0_d$ is the $d$-dimensional all 0 vector and $\mathbf I_d$ is the $d\times d$ identity matrix. The next iteration $  \Theta^{(\imath+1)}  $ is given by
\begin{equation*}
\Theta^{(\imath+1)} = \left\{
\begin{array}{ll}
\Theta^{*} & \mbox{with probability } p  \\
\Theta^{(\imath)} & \mbox{with probability } 1-p
\end{array}
\right.,
\end{equation*}
where
\begin{equation} \label{eqn:acceptence_prob}
p = \min\left( 1 \, , \, \frac{P(\Theta^*)\exp\left( - \frac{1}{2} v(\tau+\kappa \Delta\tau)^T v(\tau+\kappa \Delta\tau) \right)}{P(\Theta^{(\imath)})\exp\left( - \frac{1}{2} v(\tau)^T v(\tau) \right)} \right).
\end{equation}

We next discuss how to handle the constraints under the Bayesian framework. Under this framework, a major difference to the maximum likelihood framework lies in the fact that we need to specify prior distributions for all the unknown parameters. Those prior distributions encode our prior belief on the unknown parameters. As discussed earlier, the sign constrains on the unknowns are mathematical representations of our business knowledge regarding those parameters. Therefore, conceptually, it is natural to include sign constraints in the prior. For example, for any $i\in \mathcal H_\beta$ and any $\nu=1,\cdots,g$, let the priors of $\beta_i$ and $\beta_{i,\nu}$ be $\pi_{\beta_i}(\beta_i)$ and $\pi_{\beta_{i,\nu}}(\beta_{i,\nu})$, we make sure
$$ \pi_{\beta_i}(\omega) \, = \, 0 \,\, \forall \omega < 0, \,\, \mbox{and}\,\, \pi_{\beta_{i,\nu}}(\omega) \, = \, 0 \,\, \forall \omega < 0. $$
Similarly, we can specify priors for $\gamma_j$'s as well as $\gamma_{j,\nu}$'s for all $j \in \mathcal H_\gamma$ and $\nu = 1,\cdots,g$. We also assume that the priors of the unknowns are independent. For the simplicity of notation, we omit the subscript of the prior distribution. The joint prior distributions of all the unknown parameters is given by:
\begin{equation} \label{eqn:prior}
\begin{array}{rcl}
\pi(\Theta) &= &\displaystyle{ \left(\prod_{i=1}^{m}\left(\pi(\alpha_i)\pi(k_i)\pi(\lambda_i)\pi(\beta_i)\pi(\eta_i^2) \prod_{\nu=1}^g \pi(\beta_{i,\nu} ) \right)\right)} \\[5pt]
&& \times \displaystyle{ \left( \prod_{j=1}^{n} \left( \pi(\gamma_j)\pi(\xi_j^2) \prod_{\nu=1}^{g} \pi(\gamma_{j,\nu})\right)\right) \pi(\sigma^2)}.
\end{array}
\end{equation}
Under Bayesian framework, the posterior distribution of the parameters is proportional to the product of the prior distribution and the likelihood function. Since we have encoded the sign constraints in the prior distribution, we will not include them in the likelihood function anymore. Therefore, we have a likelihood function different from (\oldref{eqn:likelihood}) with the $\beta$ and $\gamma$ dependent on the scaling factor due to the truncation of the Normal distribution removed. We have
\begin{equation} \label{eqn:likelihood2}
\mathcal L(\Theta) \, = \, \left(\prod_{t=\ell}^{w} \prod_{\nu=1}^{g} f(y_{t, \nu}|\Theta)\right) \times
\left(\prod_{i=1}^{m}\prod_{\nu=1}^{g}f_{N}(\beta_{i, \nu} | \beta_i, \eta_i^2 )\right)
\times
\left(\prod_{j=1}^{n}\prod_{\nu=1}^{g}f_N(\gamma_{j, \nu} | \gamma_j, \xi_j^2)\right).
\end{equation}
where
\begin{equation*}
f_N(\beta_{i,\nu}|\beta_i, \eta_i^2) = \frac{1}{\sqrt{2\pi}\eta_i} \exp\left(-\frac{1}{2}\left(\frac{\beta_{i, \nu} - \beta_i}{\eta_i}\right)^2\right).
\end{equation*}
and 
\begin{equation*}
f_N(\gamma_{j,\nu}|\gamma_j, \xi_j^2) = \frac{1}{\sqrt{2\pi}\xi_j} \exp\left(-\frac{1}{2}\left(\frac{\gamma_{j, \nu} - \gamma_j}{\xi_j}\right)^2\right).
\end{equation*}
And therefore we let
$$P(\Theta) \, = \, \pi(\Theta) \mathcal L (\Theta).$$
As we can see, $P(\Theta)$ is always $0$ outside the feasible region of the optimization problem (\oldref{eqn:ML_constrained}), and therefore any proposal falls outside that the region is not accepted according to equation (\oldref{eqn:acceptence_prob}). On the other hand, $\mathcal L(\Theta)$, not involving evaluation of Normal CDF, is less complex than $L(\Theta)$ in terms of numerically evaluating its value and gradient, which is one of the most time consuming parts in the leapfrog procedure. \\

While the above treatment of the constraints guarantees that the HMC algorithm always takes legitimate samples and also mitigates the computational load, it could lead to high rejection rate, due to the lack of a ``guardrail" in the Hamiltonian dynamics to prevent an infeasible sample being proposed in the first place. We introduce a mechanism due to \citep{hmc} to provide such a guardrail in the next subsection.

\subsection{Avoiding Infeasibility in Hamiltonian Dynamics} \label{subsec:sign}
We consider box constraints on a subset of the variables, i.e., $ l_i \leq \theta_i \le u_i $ with $u_i$ can possibly be $+\infty$ and $l_i$ possibly be $-\infty$. The idea of imposing constraints is to let the potential energy be infinite for values that violate the constraints. To illustrate the idea, let $ U^{\star}(\Theta) $ be the potential energy omitting the constraints. Consistent to the constrains included in MMMs, we only consider constraints  of the format $ \theta_i \ge 0 $. Then, we have the following equation.
\begin{equation}
U(\Theta) = U^{\star}(\Theta) + C_r(\theta_i, 0),
\end{equation}
Following the formatting in \citep{hmc}, with a given $r>0$, we let
\begin{equation}
C_r(\theta_i, l_i) = \begin{cases}
0 & \text{if}~ \theta_i \ge 0 \\
r^{(r+1)}(- \theta_i)^{r} & \text{if}~ \theta_i < 0
\end{cases}.
\end{equation}
It is obvious that $  \lim\limits_{r \rightarrow +\infty} C_r(\theta_i, 0) $ is 0 for any $ \theta_i \ge 0 $ and $\infty$ for any $ \theta_i < 0 $. To simulate the dynamics based on this $ U(z) $, we can define
\begin{equation} \label{eqn:hill}
H(\Theta, v) = U^{\star}(\Theta)/2 + [ C_r(\theta_i, 0) + K(v) ] + U^{\star}(\Theta)/2.
\end{equation}
Intuitively, function $ C_r(\theta_i, 0) $ can be seen as a steep hill. The trajectory just bounces off the guard rail defined by the lower bound 0. This modification defines a variation on the leapfrog algorithm in which the half step of $v$ update remains the same, but the full step of $ \Theta $ update in (\oldref{eqn:full_step_q}) is changed. In particular, after computing
$ \theta_i^{\prime} = \theta_i(\tau) + \Delta \tau \frac{v_i(\tau+\Delta \tau/2)}{2}$, we check if $z_i^{\prime} \geq 0 $. If yes, set $ \theta_i(\tau+\Delta \tau) = \theta_i^{\prime}$ then proceed to the next steps. If not, then
$$ \theta_i(\tau+\Delta \tau) = - \theta_i^{\prime} ~~ \text{and}~~ v_i(\tau+\Delta \tau/2) = - v_i. $$

If several variables have constraints, we must follow the above procedure for each. In other words, the full step for $ \Theta $ in equation (\oldref{eqn:full_step_q}) is replaced by the proposed procedure.  \\

\section{Simulated Examples} \label{sec:exp}
In this section, we report results on simulated data for both the base model and the hierarchical model. In order to further compare the performance, in addition to the proposed marketing mix model, we also include an existing ad hoc procedure as follows. One first specifies a few candidates for the constant decreasing rate ($\alpha$) of the variable that adstocking effect needs to be considered based on experts' opinions. Then for each candidate, one computes the correlation coefficient between the transformed variable after adstocking is taken into account and the residuals of an ordinary least squares (OLS) regression of dependent variable against all independent variables excluding those with adstocking effects. The best $ \alpha $ is chosen with the largest correlation coefficient among all candidates. If there are more than one variable with adstocking effect, one has to repeat the above procedure independently for each variable to determine the best $ \alpha$. Once the process is done, one then fits a regression model with all variables to get the estimates. The whole process is ad hoc in nature as it separates the modeling procedures into two independent parts, and it also heavily depends on experts' opinions. Moreover, it does not explicitly quantify the shape and scale effects.

\subsection{Example Bundle 1 --- Base Model} \label{sec:exam1}
We first work with example bundles concerning the performance under the base model. Following the same notations, the bundle consists of the following $ 4 $ cases.
\begin{itemize}
	\item Case $ 1 $: $ m =2, w=52$
	\item Case $ 2 $: $ m =2, w=104$
	\item Case $ 3 $: $ m =4, w=104$
	\item Case $ 4 $: $ m =4, w=208$
\end{itemize}
$ n=1 $ and $ \ell =5 $ are fixed for all examples. The true model parameters are given below.
\begin{itemize}
	\item Carryover parameters $ \alpha_i = 0.5$, $ i=1, 2 $ for cases $1$ and $2$ and $ i=1,2, 3, 4 $ for cases $3$ and $4$
	\item Shape parameters $ k_i =0.2$,  $ i=1, 2 $ for cases $1$ and $2$ and $ i=1,2, 3, 4 $ for cases $3$ and $4$
	\item Scale parameters $ \lambda_i = 0.8$,  $ i=1, 2 $ for cases $1$ and $2$ and $ i=1,2, 3, 4 $ for cases $3$ and $4$
	\item Regression coefficients $ \beta $: $ \beta_i =1$,  $ i=1, 2 $ for cases $1$ and $2$ and $ i=1,2, 3, 4 $ for cases $3$ and $4$
	\item Regression coefficients $ \gamma $: $ \gamma_j=1, j=0, 1 $. Note that $ \gamma_0 = 1 $ meaning an intercept is considered in this section
	\item Variance of the residuals: $ \sigma^2 = 0.25 $
\end{itemize}
With the true parameters specified above, the prior distributions are given as follows.
\begin{itemize}
	\item Carryover parameters $ \alpha_i $, $  \alpha_i^{*} = \log(\alpha_i/(1-\alpha_i)) $ and $  \alpha_i^{*} \sim N(0, 0.5^2)$
	\item Shape parameters $ k_i $, $ k_i \sim \Gamma(0.5, 1) $
	\item Scale parameters $ \lambda_i $, $ \lambda_i \sim \Gamma(0.5, 1) $
	\item Regression coefficients $ \beta $: $ \beta_{i}, \sim TN(0, +\infty, 1, 0.5^2)  $
	\item Regression coefficients $ \gamma $: $ \gamma_{j}, \sim TN(0, +\infty, 1, 0.5^2)  $
	\item Variance of the residuals: $ \sigma^2 \sim IG(1, 1) $
\end{itemize}
IG and TN are inverse Gamma and truncated Normal distribution, respectively. We deliberately choose truncated Normal as the prior distribution for $ \beta_{i} $ and $ \gamma_j $ as a way to impose non-negative sign constraints. In addition, we apply a kernel trick on the carryover parameters, and after the logistic transformation, it works at the unbounded $\alpha_i^{*}$ scale. The scale will then be converted back to the original one before outputting the final estimates.    \\

We kicked off a run with the number of HMC iterations equaling  $ 20,000 $  and the first $ 10,000 $ iterations are treated as burn-in. The thinning parameter is fixed at $20$ after burn-in. Hence, for each unknown parameter, we obtain $ 500 $ samples in total. We also include results from two constrained optimization methods: limited memory version of bounded Broyden-Fletcher-Goldfarb-Shanno (L-BFGS-B) \citep{zhu1997algorithm} and sequential quadratic programming (SQP) \citep{boggs1995sequential} as comparisons. Each optimization method is repeated $ 20 $ times with different initial values, and the estimates with the largest log-likelihood is recorded. The implementations of these two non-linear constrained optimization methods are available in the scipy \citep{jones2001scipy} package in Python as well as in lbfgsb3 \citep{nash2015package} and NlcOptim \citep{chen2017nlcoptim} packages in R, to just name a few. For the implementation of marketing mix model with HMC, we use our self-developed Python codes. \\

\begin{figure}[htbp!]
	\centering
	\includegraphics[width=1.0\textwidth]{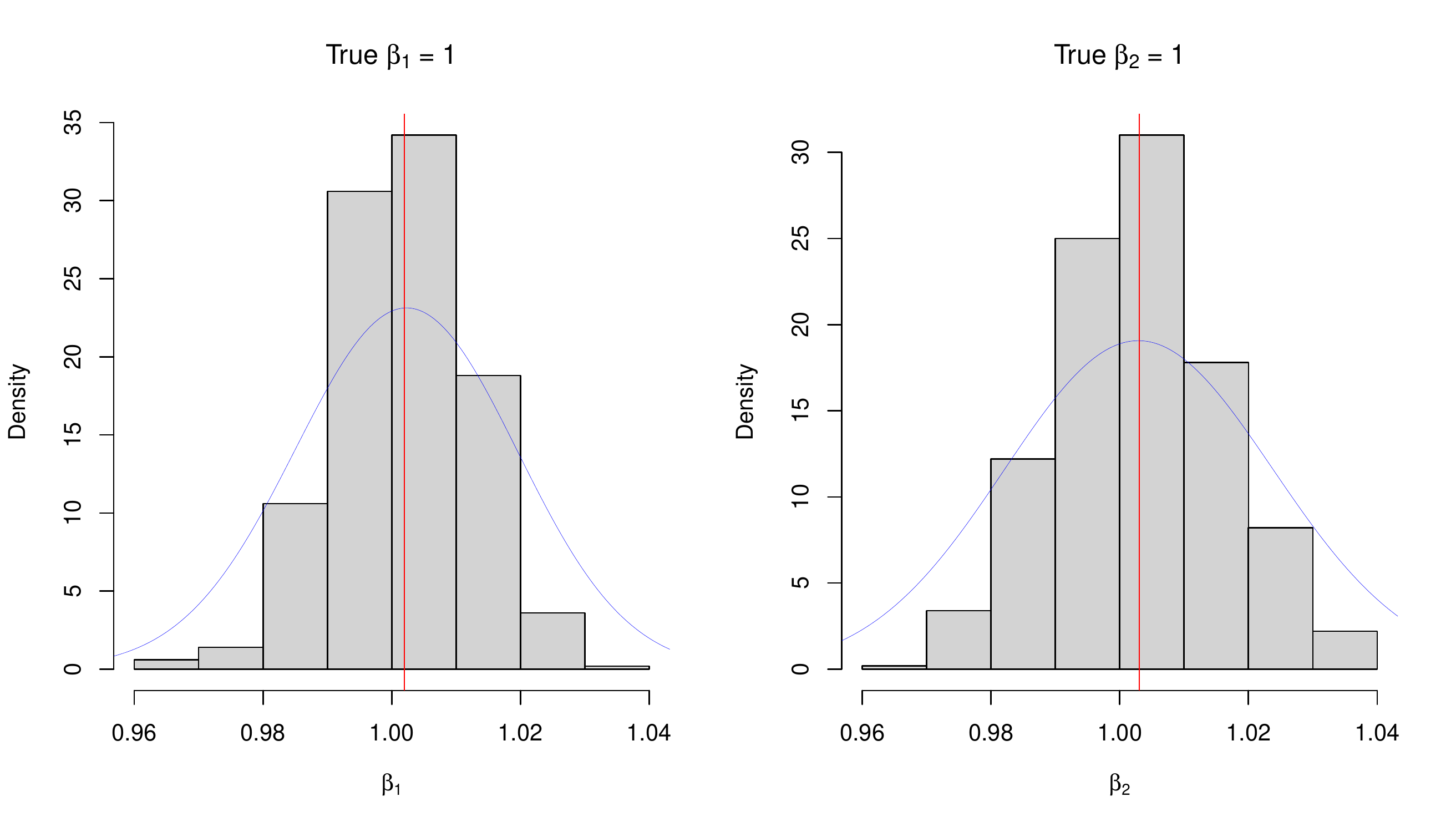}
	\caption{Histogram of regression parameters, $ \beta_1, \beta_2 $ for Case~$ 1 $ added by empirical density. The red line is the sample average. } \label{fig:case1}
\end{figure}

The root mean squared error (RMSE) of the example bundles are reported in Table~\oldref{tab:rmse_base}. The estimated parameters are reported in Tables~\oldref{tab:case1_estimate}, \oldref{tab:case2_estimate}, \oldref{tab:case3_estimate} and \oldref{tab:case4_estimate} for cases $1$, $2$, $3$ and $4$, respectively, in the Appendix. Among all the unknown parameters of the based model, we are particularly interested in the regression coefficients ($ \beta_i $), we report the histogram of $ \beta_1 $ and $ \beta_2 $ for Case~$ 1 $ in Figure~\oldref{fig:case1}. \\

From Table~\oldref{tab:rmse_base}, we observe that HMC has the smallest RMSE: it is about $ 7 $ times smaller than that of L-BFGS-B and SQP. The performance of L-BFGS-B and SQP are similar to each other, although none is able to obtain as accurate estimates as HMC does. From Figure~\oldref{fig:case1}, the histogram centers at its sample average, which is the final estimate of HMC. The density plot delivers the same message. This is anticipated as if the shape of histogram is obviously more than unimodal, it means the samples have more than one centers, and it is rather risky to use the sample mean as an estimator. Moreover, for Case~$ 2 $, we report its histogram in Figure~\oldref{fig:case2} for $ \beta_1, \beta_2 $ in the Appendix. It has a very similar pattern as observed for Case~$ 1 $. We also report the histogram of $ \beta_1, \beta_2, \beta_3 $ and $ \beta_4 $ in Figure~\oldref{fig:case3} for Case~$ 3 $. The same plot for Case~$4$ is reported in Figure~\oldref{fig:case4} in the Appendix. The performance for cases $ 3 $ and $ 4 $ are satisfactory as $ m $ advances to $ 4 $  further increasing the modeling difficulty. \\

Moreover, the ad hoc process is also considered for all of the $4$ examples. However, none is able to provide estimates that comply with the sign constraints. Taking the first two examples to illustrate the point, the estimated regression models are given in (\oldref{eqn:nosign})
\begin{eqnarray} \label{eqn:nosign}
\text{Example 1}: ~~~~ \hat{y} &=& 0.15x_1 - 0.09x_2 + 1.20z_1 + 1.57; \nonumber \\
\text{Example 2}: ~~~~ \hat{y} &=& 0.16x_1 - 0.11x_2 + 1.28z_1 + 1.54.
\end{eqnarray}
In equation (\oldref{eqn:nosign}), $ x_1, x_2, z_1 $ are associated with $ \beta_1, \beta_2, \gamma_1$ in equation (\oldref{eqn:base_model}), respectively. The estimated $ \beta $ from the ad hoc process is not directly comparable with these from the proposed marketing mix model. The ad hoc process does not have a layer of saturation so the scale of the transformed variable is different from the proposed procedure. In addition, the data is simulated from a marketing mix model, and it is, therefore, not a fair game for the ad hoc process. That being said, it is evident that the ad hoc process depends on unconstrained regression model. In practice, if any estimated parameter does not comply with its sign constraint from the ad hoc process, one has to add additional heuristics of manual adjustment, which also heavily relies on experts' opinions or one's own experience. Compared to the ad hoc process, the proposed marketing mix model handles sign constraints automatically ``on the fly". It not only streamlines the whole modeling process, but also tremendously reduces the dependency of experts' opinions. \\

\begin{table}
	\centering
	\caption{RMSE of HMC, L-BFGS-B and SQP for the Base Model} \label{tab:rmse_base}
	\begin{tabular}{|c|c|c|c|}
		\hline
		& HMC & L-BFGS-B & SQP   \\ \hline
		Case $1$ & $0.064$ & $0.477$ & $0.536$   \\ \hline
		Case $2$ & $0.082$ & $0.435$ & $0.213$   \\ \hline
		Case $3$ & $0.049$ & $0.394$ & $0.417$   \\ \hline
		Case $4$ & $0.055$ & $0.344$ & $0.534$   \\ \hline
	\end{tabular}
\end{table}

\begin{figure}[htbp!]
	\hspace{-10mm}%
	\includegraphics[width=1.2\textwidth]{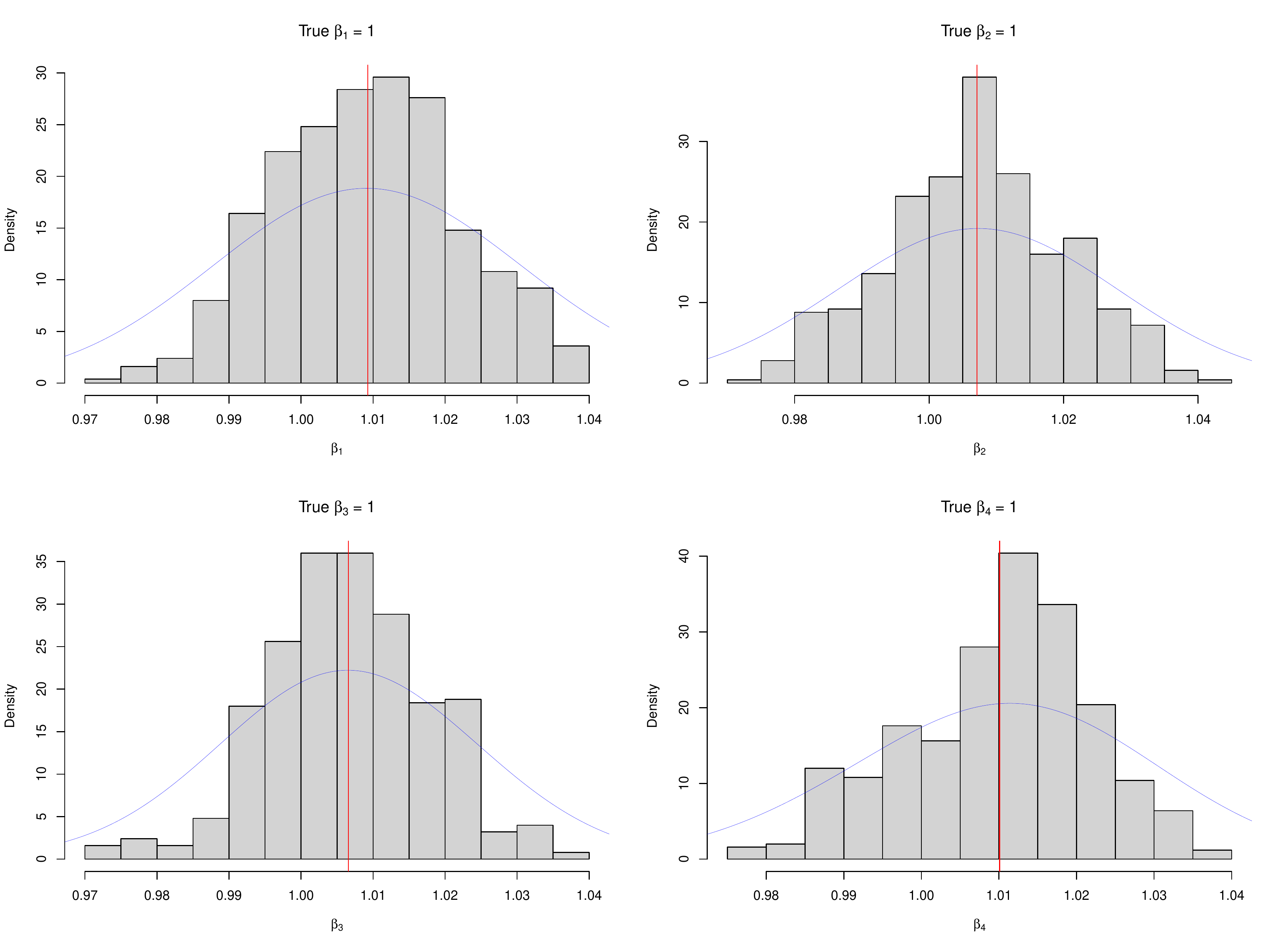}
	\caption{Histogram of regression parameters, $ \beta_1, \beta_2, \beta_3, \beta_4 $ for Case~$ 3 $ added by empirical density. The red line is the sample average.} \label{fig:case3}
\end{figure}

\subsection{Example Bundle 2 --- Hierarchical Model}
\begin{figure}[htbp!]
	\centering
	\includegraphics[width=1.0\textwidth]{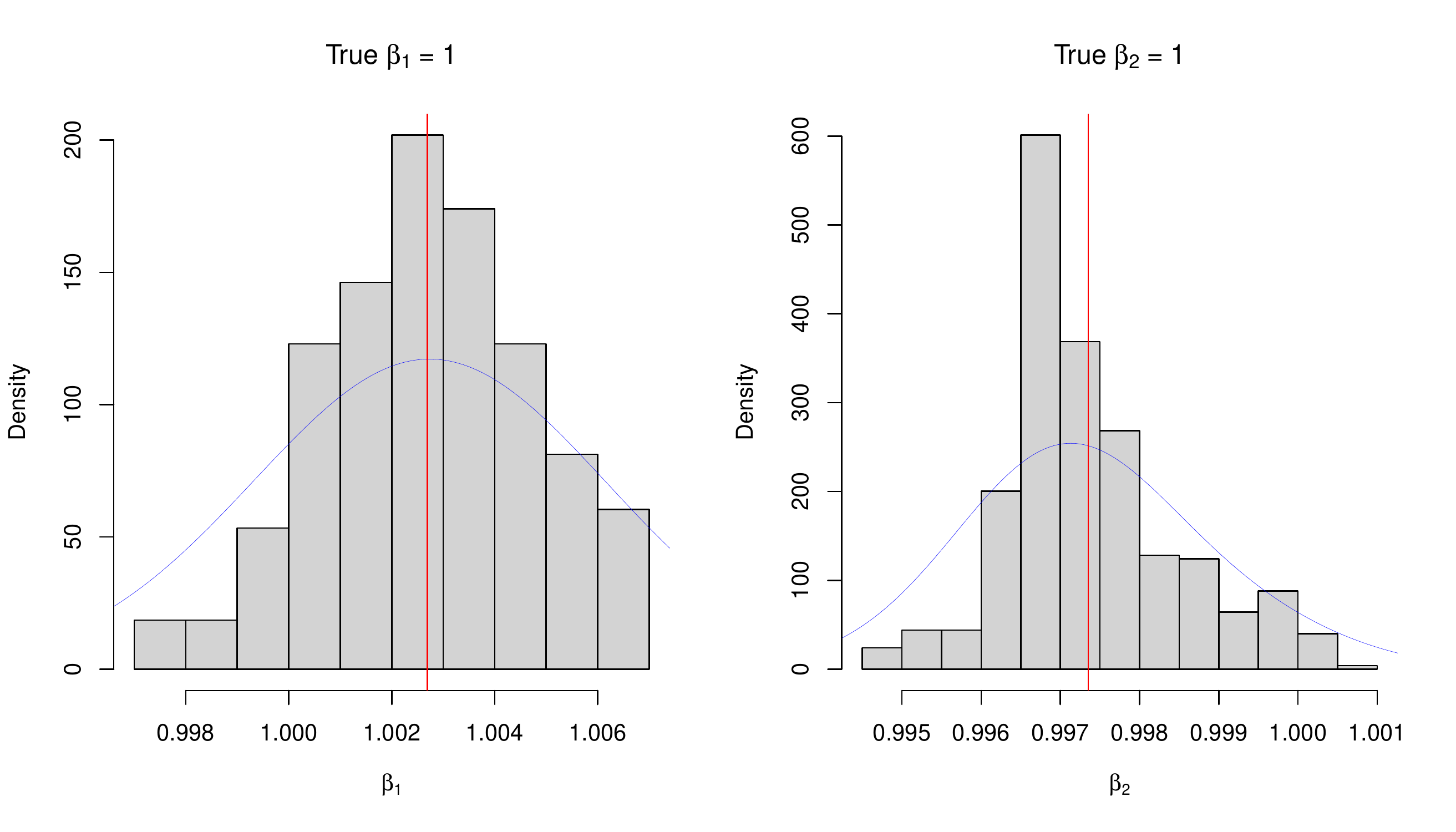}
	\caption{Histogram of fixed effects, $ \beta_1, \beta_2 $ for Case~$ 5 $ added by empirical density. The red line is the sample average. } \label{fig:case5_fe}
\end{figure}

\begin{figure}[htbp!]
	\hspace{-10mm}%
	\includegraphics[width=1.2\textwidth]{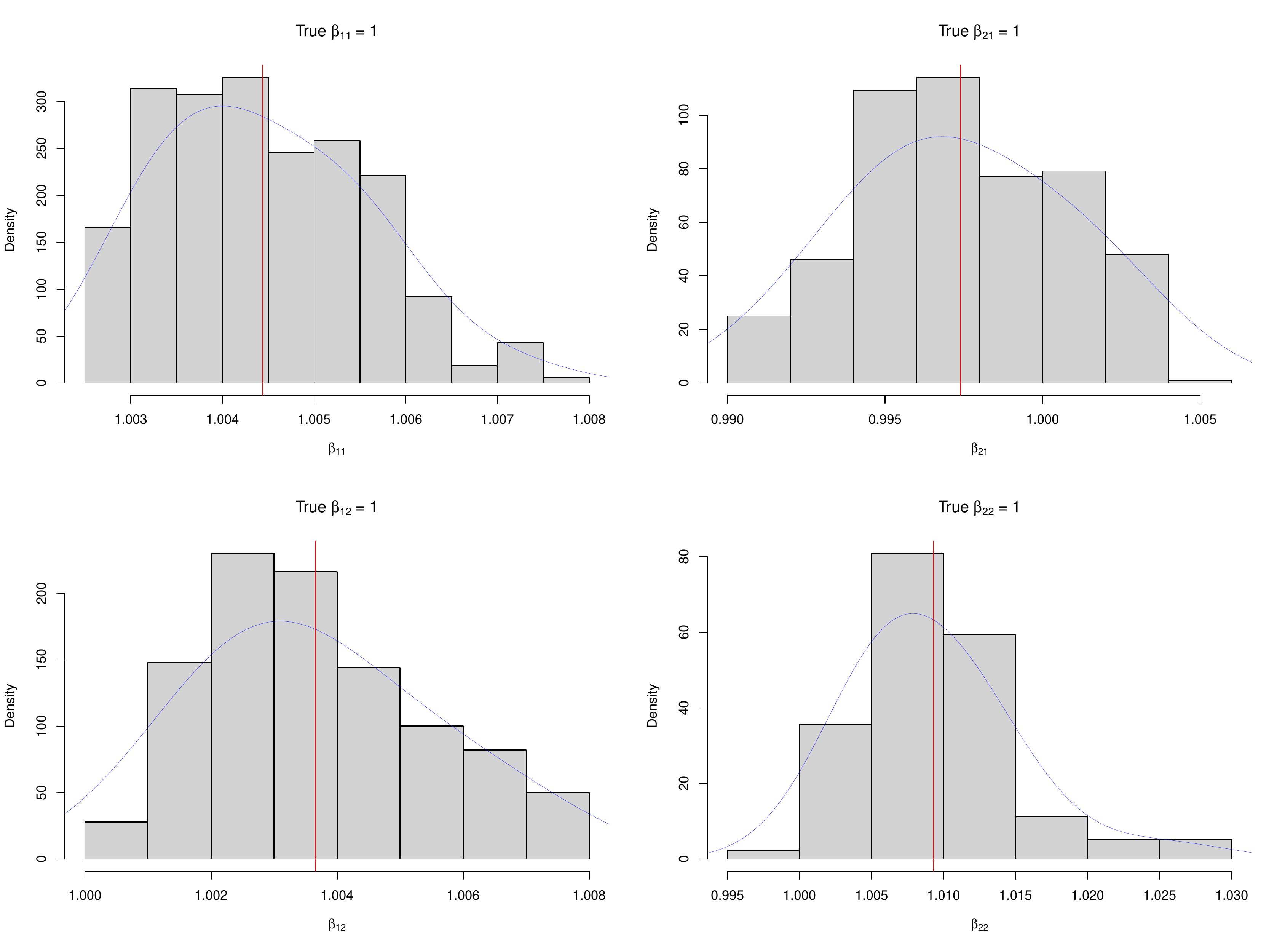}
	\caption{Histogram of random effects, $ \beta_{1, 1}, \beta_{2, 1}, \beta_{1, 2}, \beta_{2, 2} $ for Case~$ 5 $ added by empirical density. The red line is the sample average. } \label{fig:case5_rf}
\end{figure}

In this part, we assess the performance of the three methods under the hierarchical model. The proposed example bundle includes $4$ examples below.
\begin{itemize}
	\item Case $ 5 $: $ m =2, w=52, g=2$
	\item Case $ 6 $: $ m =2, w=104, g=2$
	\item Case $ 7 $: $ m =4, w=104, g=2$
	\item Case $ 8 $: $ m =4, w=208, g=2$
\end{itemize}
$ n=1 $ and $ \ell =5 $ are fixed for all examples as before. All of the true parameters are same as in Section \oldref{sec:exam1} except as $ g=2 $, we will need to consider both the fixed effects and random effects for regression parameters. The true model parameters and priors are same as the ones used for the base model except the fixed effects and random effects are new under the hierarchical model, and they are given below.
\begin{itemize}
	\item Fixed means $ \beta_i $: $ \beta_{i} = 1$, $ i=1, 2 $ for cases $5$ and $6$ and $ i=1,2, 3, 4 $ for cases $7$ and $8$
	\item Fixed means $ \gamma_j $: $ \gamma_{j} = 1 $, $ j=1, 2 $ for cases $5$ and $6$ and $ j=1,2, 3, 4 $ for cases $7$ and $8$
	\item Fixed variances $ \eta^2_i $: $ \eta_{i}^2 = 0.25$, $ i=1, 2 $ for cases $5$ and $6$ and $ i=1,2, 3, 4 $ for cases $7$ and $8$
	\item Fixed variances $ \xi^2_j $: $ \xi_{j}^2 = 0.25$, $ j=1, 2 $ for cases $5$ and $6$ and $ j=1,2, 3, 4 $ for cases $7$ and $8$
	\item Random coefficients $ \beta_{i, g} $, $\beta_{i, g} = 1$, $ i=1, 2 $ for cases $5$ and $6$ and $ i=1,2, 3, 4 $ for cases $7$ and $8$, $ g=1, 2 $
	\item Random coefficients $ \gamma_{j, g} $, $\gamma_{j, g} = 1$, $ j=1, 2 $ for cases $5$ and $6$ and $ j=1,2, 3, 4 $ for cases $7$ and $8$, $ g=1, 2 $    	
\end{itemize}
The priors are:
\begin{itemize}
	\item Fixed means, $ \beta $  $ \beta_{i}, \sim TN(0, +\infty, 1.0, 0.5^2)  $
	\item Fixed means, $ \gamma $ $ \gamma_{j}, \sim TN(0, +\infty, 1.0, 0.5^2)  $
	\item Fixed variances $ \eta^2 $: $ \eta_{i}^2, \sim TN(0, +\infty, 1.0, 0.5^2)  $
	\item Fixed variances $ \xi^2 $: $ \xi_{j}^2, \sim TN(0, +\infty, 1.0, 0.5^2)  $
	\item Random coefficients $ \beta_{i, g} \sim \sim TN(0, +\infty, 1.0, 0.5^2) $
	\item Random coefficients $ \gamma_{j, g} \sim \sim TN(0, +\infty, 1.0, 0.5^2) $
\end{itemize}
With the same settings as those in Section \oldref{sec:exam1}, we report the RMSE of the three methods in Table~\oldref{tab:rmse_hier}. The estimated parameters are reported in Tables~\oldref{tab:case5_estimate}, \oldref{tab:case6_estimate}, \oldref{tab:case7_estimate}, \oldref{tab:case8_estimate} for cases $5$, $6$, $7$ and $8$, respectively, in the Appendix. Among all the unknown parameters of the hierarchical model, we are particularly interested in the fixed means ($ \beta_i $), we report the histogram of $ \beta_1, \beta_2 $ for Case~$ 5 $ in Figure~\oldref{fig:case5_fe} for HMC. The histogram for estimated random coefficients $ \beta_{i, g}  $ of Case~$ 5 $ are also reported in Figure~\oldref{fig:case5_rf}. \\

Table~\oldref{tab:rmse_hier} is consistent with Table~\oldref{tab:rmse_base} that the HMC has the smallest RMSE value, indicating its superior performance over the other two optimization methods. The fact that HMC performs well can also be shown by the histogram and density plot of its fixed means and random coefficients reported in Figures \oldref{fig:case5_fe} and \oldref{fig:case5_rf}. In a similar pattern, we report the same plots for Case~$ 6 $ in Figures~\oldref{fig:case6_fe} and \oldref{fig:case6_rf} in the Appendix. As for cases $ 7 $ and $ 8 $, $ m $ has increased to $ 4 $ which further inflates the parameter space. For these two examples, we report the histogram of its fixed means in Figure~\oldref{fig:case7} for Case~$ 7 $, and in Figure~\oldref{fig:case8}  for Case~$ 8 $ in the Appendix. The conclusion we draw from the base model also hold here for the hierarchical model: comparing to the other two optimization methods, HMC has a best ability of recovering the ``true" parameters in simulated examples. \\

In addition, the ad hoc process is also considered for all of the 4 examples in this Section. However, none is able to provide estimate that comply with the sign constraints. The observation is consistent with that made in Section \oldref{sec:exam1}. Therefore, results from the ad hoc process have been excluded from discussion.   	

\begin{table}
	\centering
	\caption{RMSE of HMC, L-BFGS-B and SQP for the Hierarchical Model} \label{tab:rmse_hier}
	\begin{tabular}{|c|c|c|c|}
		\hline
		& HMC & L-BFGS-B & SQP   \\ \hline
		Case $5$ & $0.014$ & $0.538$ & $0.672$   \\ \hline
		Case $6$ & $0.013$ & $0.398$ & $0.656$   \\ \hline
		Case $7$ & $0.010$ & $0.489$ & $0.649$   \\ \hline
		Case $8$ & $0.007$ & $0.452$ & $0.387$   \\ \hline
	\end{tabular}
\end{table}

\begin{figure}[htbp!]
	\hspace{-10mm}%
	\includegraphics[width=1.2\textwidth]{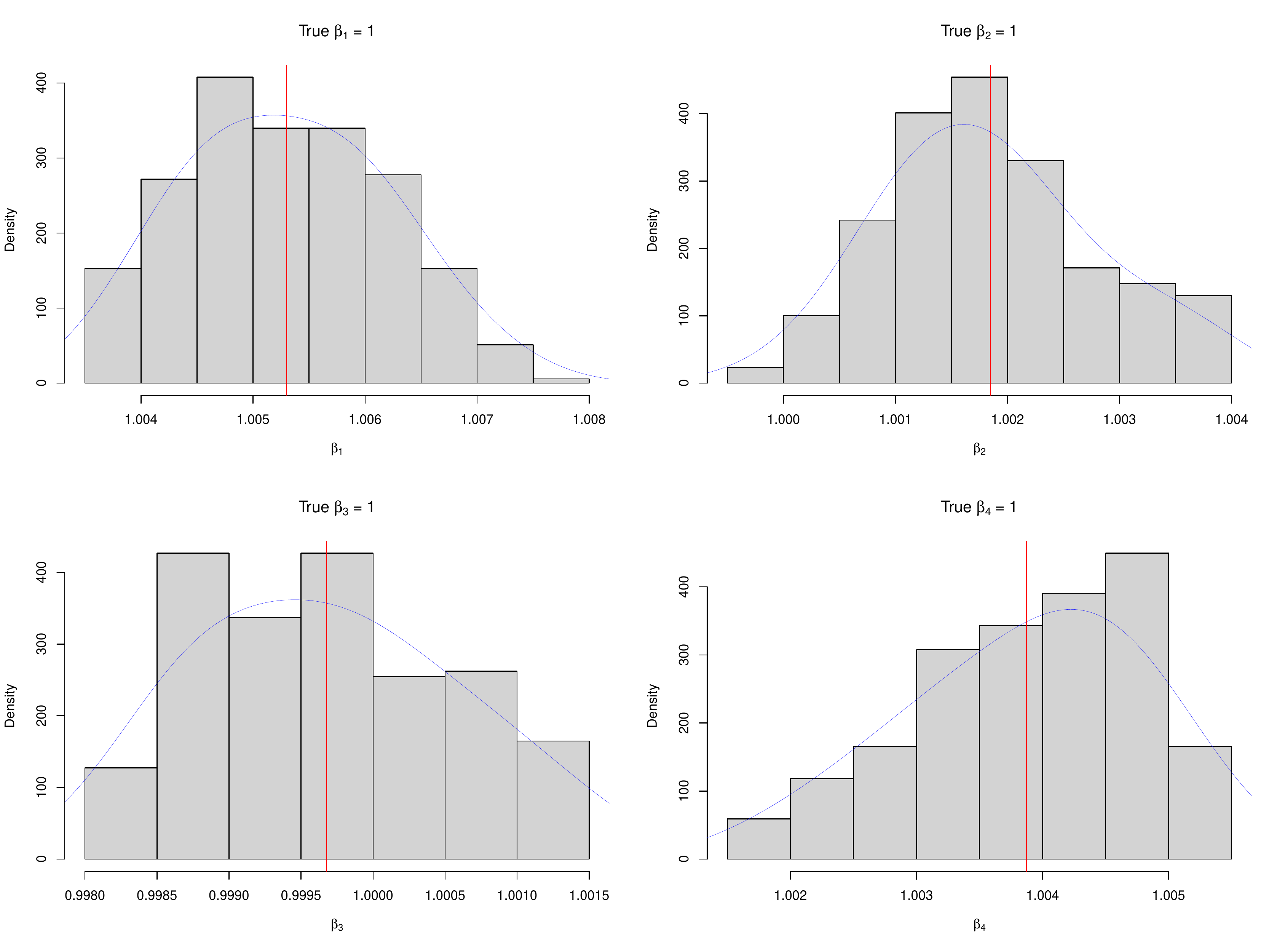}
	\caption{Histogram of regression parameters, $ \beta_1, \beta_2, \beta_3, \beta_4 $ for Case~$ 7 $ added by empirical density. The red line is the sample average.} \label{fig:case7}
\end{figure}

\newpage
\section{Analysis on Real Datasets} \label{sec:real}
In this section, we consider two real world applications. The plan is as follows. In the first example, we will mainly compare the model performance between the ad hoc process and the proposed marketing mix model. In the second example, a deeper analysis is provided using the proposed model with a test of hypothesis of critical variables.

\subsection{Real Application 1} \label{sec:realexample1}

In this example, data has been collected from a clothing retailer that contains weekly sales information for the most recent $104$ weeks. The descriptions of variables are given in Table \oldref{tab:g4}.

\begin{table}[H]
	\caption{Descriptions of variables for the real dataset.} \label{tab:g4}
	\begin{center}
	\resizebox{\textwidth}{!}{%
    \begin{tabular}{|l|c|c|c|c|} \hline
		\textit{Variable} & \textit{Adstocking Effect} & \textit{Ind Var} & \textit{Dep Var} & \textit{Pos Sign Cons} \\ \hline
		Television TRP  & Yes & Yes & No & Yes \\ \hline
		Outdoor Impression  & Yes & Yes & No & Yes \\ \hline \hline
		Catalina Coupon Distributed Quantity & No & Yes & No & Yes \\ \hline
		Digital Marketing Distributed Quantity & No & Yes & No & Yes \\ \hline
		Digital Display Impression   & No & Yes & No & Yes \\ \hline
		Digital Facebook Impression & No & Yes & No & Yes \\ \hline
		Digital Instagram Impression & No & Yes & No & Yes \\ \hline
		Digital Pinterest Impression & No & Yes & No & Yes \\ \hline
		Digital Paidsearch Impression & No & Yes & No & Yes \\ \hline
		Digital Youtube Impression & No & Yes & No & Yes \\ \hline
		Seasonality Index & No & Yes & No & No \\ \hline
		Unemployment Rate & No & Yes & No & No \\ \hline \hline
		Natural Logarithm of Sales Quantity & No & No & Yes & N.A. \\ \hline
	\end{tabular}}
	\end{center}
\end{table}

Abbreviated forms are used to save space in Table \oldref{tab:g4} as follows: \textit{Ind Var} is independent variable; \textit{Dep Var} is dependent variable; \textit{Pos Sign Cons} is positive sign constraint. targeted rating point (TRP) is computed as the percentage of the target audience reached by an advertisement through a medium, for example, if there are $1,000$ impressions among the $10,000$ target audience, the TRPs is $(1,000/10,000) \times 100 = 10$. Impressions are realized when an advertisement or any other form of digital media is displayed on an user's device. Impressions are not action-based and are defined by a user potentially seeing the advertisement. For example, for $ 10 $ users each views an advertisement $5$ times in a given week. The impressions of the advertisement for that week are $ 10\times 5 = 50 $. Essentially, together with self-explanatory distributed quantity, all of these variables are common metrics used to quantify the execution of marketing campaigns. In addition, the data have been further aggregated to store cluster level to avoid missing value issues as much as possible, and $4$ store clusters are considered in this example. \\

We consider both the marketing mix model and the ad hoc process described in Section \oldref{sec:exp}. The marketing mix model will be estimated by HMC since it has better performance than the other two optimization methods shown in the simulation study. The model performance is measured by the marginal $ R^2 $ and conditional $ R^2 $ (\cite{nakagawa2013general}). Both metrics are standard measures of model fitting in a linear mixed effects model. With a slight abuse of notation, we write the formula in the following equations.
\begin{equation}
\text{mar}_{R^2} = \frac{\sum_{t=\ell}^{N}(\hat{y}_t - \bar{y})^2/N}{\sum_{t=\ell}^{N}(\hat{y}_t - \bar{y})^2/N + \sum_{i=1}^{m} \eta_i^2 + \sum_{j=1}^{n} \xi_j^2 + \sigma^2}
\end{equation}

\begin{equation}
\text{con}_{R^2} = \frac{\sum_{t=\ell}^{N}(\hat{y}_t - \bar{y})^2/N + \sum_{i=1}^{m} \eta_i^2 + \sum_{j=1}^{n} \xi_j^2}{\sum_{t=\ell}^{N}(\hat{y}_t - \bar{y})^2/N + \sum_{i=1}^{m} \eta_i^2 + \sum_{j=1}^{n} \xi_j^2 + \sigma^2}
\end{equation}
The marginal $ R^2 $ measures the percentage of variance that the fixed effects can explain: the numerator is the variance of fixed effects, while the denominator is the total variance of the model: variance of fixed effects, variance of all random effects and variance of the error. In a similar fashion, conditional $R^2$ depicts the percentage of the variance that the whole regression model, i.e, both the fixed effects and random effects can explain. These two metrics are natural extensions of the usual $ R^2 $ to mixed effects models. \\

The model performance is reported in Table \oldref{tab:g5}. For both metrics, the proposed model is better. In addition, we also observe that some of the estimates of the ad hoc process do not comply with the sign constraints, which agrees with the observations made in Section \oldref{sec:exp}. In practice, additional heuristics will be employed to adjust the input dataset and/or arbitrarily ``correct'' the estimated parameters, which leads to further deterioration of performance. With such disadvantages in the ad hoc process, the proposed model provides an attractive alternative to practitioners.

\begin{table}[H]
	\caption{Model Performance of Real Application $1$} \label{tab:g5}
	\centering
	\begin{tabular}{|l|c|c|}
		\hline
		& Marginal $R^2$  & Conditional $R^2$  \\ \hline
		ad hoc process & 41.5\% & 65.8\% \\ \hline
		marketing mix model & 47.8\% & 69.7\%  \\ \hline
	\end{tabular}
\end{table}

\newpage
\subsection{Real Application 2}  \label{sec:realexample2}

We consider another dataset from an actual TV marketing campaign. It contains sales information on four different stores, and each store has two years of weekly sales quantity data on a product. To be more specific, the dependent variable is weekly natural logarithm of sales quantity of that product. There are seven independent variables. The first four variables are in units of TRPs representing advertising channels $1$, $2$, $3$ and $4$. The remaining three variables are nuisance variables: regular price, discounted price and seasonality, which are critically influence sales quantity, but are of no research interest in themselves with respect to marketing effectiveness. \\

\begin{figure}[htbp!]
	\hspace{-10mm}%
	\includegraphics[width=1.2\textwidth]{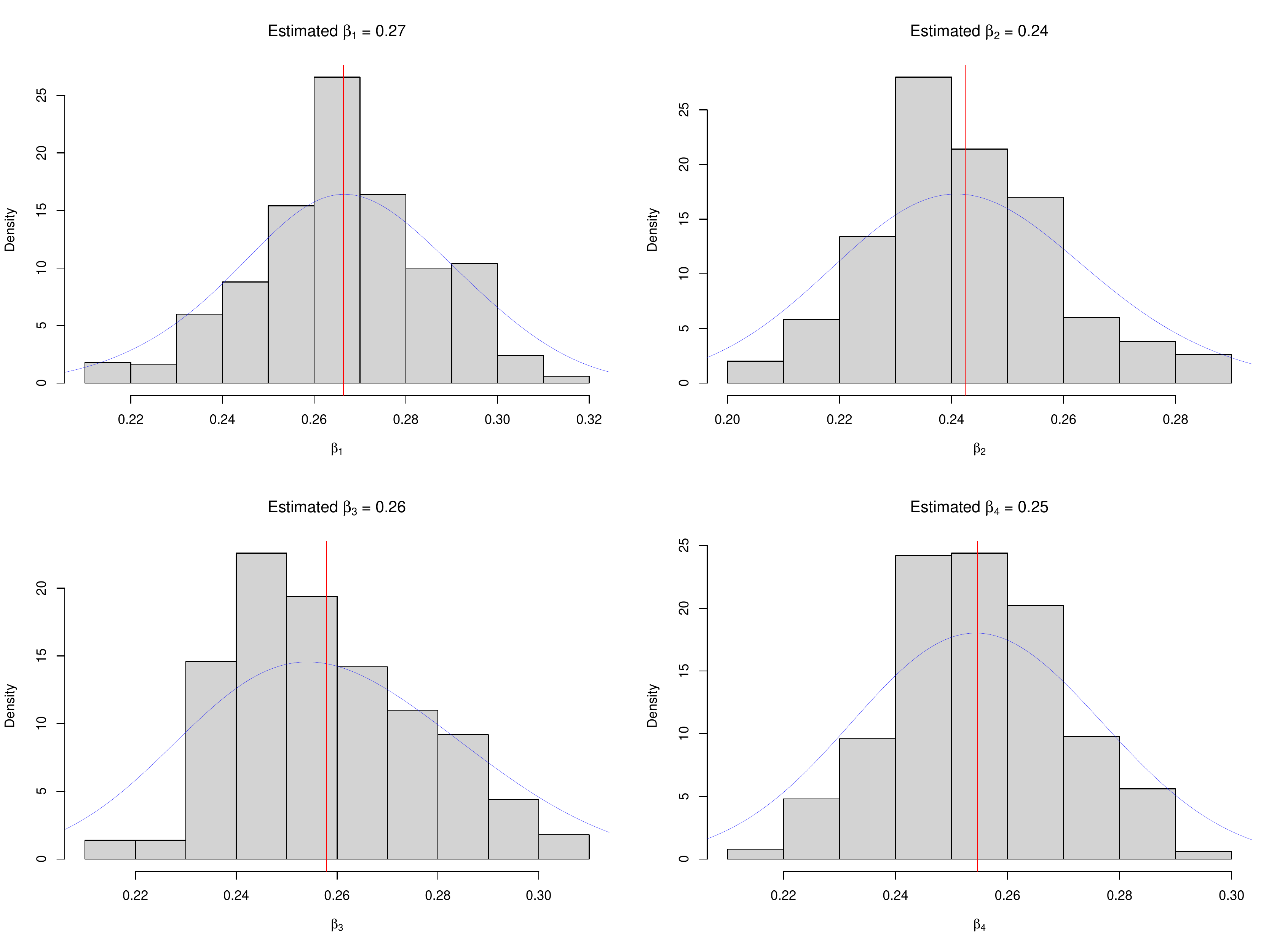}
	\caption{Histogram of regression parameters, $ \beta_1, \beta_2, \beta_3, \beta_4 $ for the Application added by empirical density. The red line is the sample average.} \label{fig:app}
\end{figure}

Following the same notations, we have $ m=4, n=3, w=104, \ell =5, g=4$. We consider the proposed marketing mix model with HMC only in this section. With the same settings as in the simulation section, we reported the estimates of the unknown parameters in Table~\oldref{tab:g3} in the Appendix to save space. We are particularly interested in the fixed regression means ($ \beta_1, \beta_2, \beta_3, \beta_4 $), which measures the group effects of how the sales changes as one unit of TRP increases. They are also constrained as non-negative as it is believed that an advertisement will at least not decrease sales. The histogram of the $ 4 $ fixed regression means are reported in Figure~\oldref{fig:app}. In addition, we are also interested in conducting a test of hypothesis that $ H_0: \beta_i = 0  $ v.s. $ H_1: \beta_i > 0  $, for $ i=1, 2, 3, 4 $ under the significance level $0.05$. Since $ 500 $ samples are collected from the posterior distribution, we also report the empirical $2.5\%$ and $ 97.5\% $ quantiles as the lower bound and upper bound for the $ 4 $ regression means in Table~\oldref{tab:inf}.\\

\begin{table}
	\caption {The $4$ regression parameters from HMC and their empirical lower and upper bounds for the real data analysis}  \label{tab:inf}
	\centering
	\begin{tabular}{|c|c|c|c|}
		\hline
		& Estimate & 2.5\% Lower Bound & 97.5\% Upper Bound \\ \hline
		$\beta_1$ & 0.27 & 0.227 & 0.300 \\ \hline
		$\beta_2$ & 0.24 & 0.212 & 0.281 \\ \hline
		$\beta_3$ & 0.26 & 0.230 & 0.297 \\ \hline
		$\beta_4$ & 0.25 & 0.224 & 0.286 \\ \hline
	\end{tabular}
\end{table}

From Table~\oldref{tab:g3} in the Appendix, we observe that compared to the simulation, the random coefficients fluctuate more around their fixed means in the real example. This was confirmed by the large variances of the random effects, for example $ \eta_{1}^2 = 0.25 $ whose magnitude is as big as the fixed mean $ \beta_1 $. This observation further supports the use of hierarchical effects model in practice, where the individual regression parameters are allowed to vary around the fixed mean, and the volatility is explicitly quantified by random effects variances, which are also treated as unknown parameters. In addition, we also observe all of the $ 4 $ regression parameters are significantly greater than $ 0 $ under significance level $0.05$, since both the lower bounds and upper bounds of its empirical $95\%$ credible intervals are above $0$. The histograms in Figure~\oldref{fig:app} looks satisfactory for a real world example.

\section{Concluding Remarks} \label{sec:con}
Marketing mix models have been widely used among practitioners as a standard way to quantify the effectiveness of advertising activities. However, the process is largely ad hoc and some parameters are set based on experience rather than derived from the data itself. In this research, we attempt to reduce ad hoc influence as much as possible by systematizing the whole process and making it more data-driven: we introduce nonlinear functions with unknown parameters to capture the carryover, shape and scale effects. In addition, we propose two models: the first is the base model where only the fixed effects are considered. The second model is one with hierarchical effects utilizing both the fixed effects and random effects to counter heterogeneity. All of the unknown parameters are simultaneously learned by both HMC, which is a novel Bayesian method originated from the study of Hamiltonian dynamics in physics and by two optimization methods. Moreover, sign constraints are also taken into account via proper specification of prior distributions as well as the enhancement to the Leapfrog algorithm discussed in Section \oldref{subsec:sign}. With the sign constraints encoding natural outcome of marketing activities, the resulting marketing mix models are more realistic from the perspective of practitioners. \\

The proposed marketing mix model represents an attractive alternative over the ad hoc process described in Section \oldref{sec:exp}. It not only streamlines the modeling process as an entire entity, but also incorporates the sign constraints automatically through the model specification. The ad hoc process is considered for all the examples in Section \oldref{sec:exp} as well as the real application in Section \oldref{sec:realexample1}, but it is unable to provide estimates that comply with all sign constraints in any of the examples. In practice, heuristics to correct the signs of the parameters will be employed further introducing subjectiveness in the measures. \\

The superior performance of HMC over the other two optimization methods that we observed in Section~\ref{sec:exp} confirms the usefulness of Bayesian method especially when the dimension is high. By using a Bayesian method, it is much easier to conduct a test of hypothesis on the regression parameters as we have seen in the real application in Section \oldref{sec:realexample2}: confidence/credible intervals are straightforward to construct when samples are collected from its posterior distribution using HMC. This is a clear advantage to methods based on the frequentist paradigm, where one usually has to rely on the asymptotic distribution of its estimator for statistical inference for most non-trivial examples. \\

Admittedly, there are still some areas where we can continue to address. For example, we assume a constant carryover effect for each advertisement. However, a more sophisticated way of quantifying the carryover, shape and scale effects is that one could allow all of the three effects vary across different regions, although it might dramatically inflate the number of unknown parameters. All in all, we believe that systematizing and standardizing marketing mix model, and letting the data speak through the model is crucial to the success of any marketing analytics application, especially in this big data era. \\



\bibliographystyle{elsarticle-harv}
\bibliography{sample}


\section*{About the Authors}
Hao Chen received his Ph.D. in Statistics from the University of British Columbia, and is currently a Senior Data Scientist at Precima. \\

Minguang Zhang holds a MSc in Economics from the Northern Illinois University, and is currently an Associate Director of Research and Development at Precima. \\

Lanshan Han holds a Ph.D. in Decision Sciences and Engineering Systems from the Rensselaer Polytechnic Institute, and is currently a Director of Research and Development at Precima. \\

Alvin Lim received his Ph.D. in Mathematical Sciences from the Johns Hopkins University, and is currently Precima’s Chief Scientist and Vice President for Research and Development.








\newpage
\appendix
\section{Additional Tables and Figures in Sections 5 and 6}

\begin{table}[htbp!]
	\centering
	\caption{Case $1$: Estimated Parameters} \label{tab:case1_estimate}
	\begin{tabular}{|c|c|c|c|c|}  \hline
		& TRUE & HMC & L-BFGS-B & SQP \\ \hline
		$\alpha_1$ & 0.5 & 0.498 & 0.112 & 0.333 \\ \hline
		$\alpha_2$ & 0.5 & 0.509 & 0.851 & 0.501 \\ \hline
		$k1$ & 0.2 & 0.268 & 0.956 & 0.522 \\ \hline
		$k2$ & 0.2 & 0.118 & 0.528 & 0.391 \\ \hline
		$\lambda_1$ & 0.8 & 0.623 & 0.605 & 0.007 \\ \hline
		$\lambda_2$ & 0.8 & 0.758 & 0.312 & 0.14 \\ \hline
		$\beta_1$ & 1.0 & 1.002 & 1.263 & 0.768 \\ \hline
		$\beta_2$ & 1.0 & 1.003 & 1.498 & 0.49 \\ \hline
		$\gamma_0$ & 1.0 & 1.006 & 0.918 & 0.49 \\ \hline
		$\gamma_1$ & 1.0 & 1.005 & 0.023 & 2.161 \\ \hline
		$\sigma^2$ & 0.25 & 0.24 & 0.242 & 0.23 \\ \hline
	\end{tabular}
\end{table}

\begin{table}[htbp!]
	\centering
	\caption{Case $2$: Estimated Parameters} \label{tab:case2_estimate}
	\begin{tabular}{|c|c|c|c|c|}
		\hline
		& TRUE & HMC & L-BFGS-B & SQP \\ \hline
		$\alpha_1$ & 0.5 & 0.488 & 0.589 & 0.692 \\ \hline
		$\alpha_2$ & 0.5 & 0.474 & 0.85 & 0.656 \\ \hline
		$k_1$ & 0.2 & 0.365 & 0.936 & 0.72 \\ \hline
		$k_2$ & 0.2 & 0.135 & 0.522 & 0.305 \\ \hline
		$\lambda_1$ & 0.8 & 0.598 & 0.719 & 0.84 \\ \hline
		$\lambda_2$ & 0.8 & 0.775 & 0.099 & 0.715 \\ \hline
		$\beta_1$ & 1.0 & 1.002 & 0.722 & 0.958 \\ \hline
		$\beta_2$ & 1.0 & 1.002 & 0.801 & 0.737 \\ \hline
		$\gamma_0$ & 1.0 & 1.001 & 1.707 & 0.896 \\ \hline
		$\gamma_1$ & 1.0 & 1.001 & 0.567 & 1.257 \\ \hline
		$\sigma^2$ & 0.25 & 0.232 & 0.234 & 0.225 \\ \hline
	\end{tabular}
\end{table}

\begin{table}[htbp!]
	\centering
	\caption{Case $3$: Estimated Parameters} \label{tab:case3_estimate}
	\begin{tabular}{|c|c|c|c|c|}
		\hline
		& TRUE & HMC & L-BFGS-B & SQP \\ \hline
		$\alpha_1$ & 0.5 & 0.507 & 0.343 & 0.596 \\ \hline
		$\alpha_2$ & 0.5 & 0.514 & 0.398 & 0.41 \\ \hline
		$\alpha_3$ & 0.5 & 0.51 & 0.606 & 0.67 \\ \hline
		$\alpha_4$ & 0.5 & 0.501 & 0.965 & 0.686 \\ \hline
		$k_1$ & 0.2 & 0.078 & 0.93 & 0.552 \\ \hline
		$k_2$ & 0.2 & 0.311 & 0.771 & 0.096 \\ \hline
		$k_3$ & 0.2 & 0.227 & 0.362 & 0.069 \\ \hline
		$k_4$ & 0.2 & 0.312 & 0.307 & 0.308 \\ \hline
		$\lambda_1$ & 0.8 & 0.837 & 0.609 & 0.368 \\ \hline
		$\lambda_2$ & 0.8 & 0.786 & 0.493 & 0.269 \\ \hline
		$\lambda_3$ & 0.8 & 0.784 & 0.062 & 0.638 \\ \hline
		$\lambda_4$ & 0.8 & 0.756 & 1 & 0.007 \\ \hline
		$\beta_1$ & 1.0 & 1.009 & 1.521 & 1.126 \\ \hline
		$\beta_2$ & 1.0 & 1.007 & 0.201 & 1.134 \\ \hline
		$\beta_3$ & 1.0 & 1.007 & 1.052 & 2.212 \\ \hline
		$\beta_4$ & 1.0 & 1.01 & 1.236 & 0.5 \\ \hline
		$\gamma_0$ & 1.0 & 1.005 & 0.67 & 0.607 \\ \hline
		$\gamma_1$ & 1.0 & 1.004 & 1.045 & 0.823 \\ \hline
		$\sigma^2$ & 0.25 & 0.237 & 0.238 & 0.256 \\ \hline
	\end{tabular}
\end{table}

\begin{table}[htbp!]
	\centering
	\caption{Case $4$: Estimated Parameters} \label{tab:case4_estimate}
	\begin{tabular}{|c|c|c|c|c|}
		\hline
		& TRUE & HMC & L-BFGS-B & SQP \\ \hline
		$\alpha_1$ & 0.5 & 0.505 & 0.703 & 0.898 \\ \hline
		$\alpha_2$ & 0.5 & 0.504 & 0.269 & 0.372 \\ \hline
		$\alpha_3$ & 0.5 & 0.494 & 0.178 & 0.941 \\ \hline
		$\alpha_4$ & 0.5 & 0.481 & 0.864 & 0.811 \\ \hline
		$k_1$ & 0.2 & 0.341 & 0.087 & 0.097 \\ \hline
		$k_2$ & 0.2 & 0.311 & 0.669 & 0.952 \\ \hline
		$k_3$ & 0.2 & 0.296 & 0.201 & 0.605 \\ \hline
		$k_4$ & 0.2 & 0.091 & 0.371 & 0.720 \\ \hline
		$\lambda_1$ & 0.8 & 0.763 & 0.859 & 0.253 \\ \hline
		$\lambda_2$ & 0.8 & 0.791 & 0.911 & 0.079 \\ \hline
		$\lambda_3$ & 0.8 & 0.756 & 0.364 & 0.584 \\ \hline
		$\lambda_4$ & 0.8 & 0.783 & 0.364 & 0.693 \\ \hline
		$\beta_1$ & 1.0 & 0.996 & 0.372 & 2.187 \\ \hline
		$\beta_2$ & 1.0 & 0.985 & 1.269 & 1.774 \\ \hline
		$\beta_3$ & 1.0 & 0.996 & 1.340 & 0.662 \\ \hline
		$\beta_4$ & 1.0 & 1.001 & 0.910 & 0.657 \\ \hline
		$ \gamma_0 $ & 1.0 & 0.997 & 1.526 & 0.223 \\ \hline
		$ \gamma_1 $ & 1.0 & 0.992 & 0.370 & 0.518 \\ \hline
		$ \sigma^2 $ & 0.25 & 0.244 & 0.246 & 0.240 \\ \hline
	\end{tabular}
\end{table}

\begin{figure}[htbp!]
	\hspace{-10mm}%
	\includegraphics[width=1.2\textwidth]{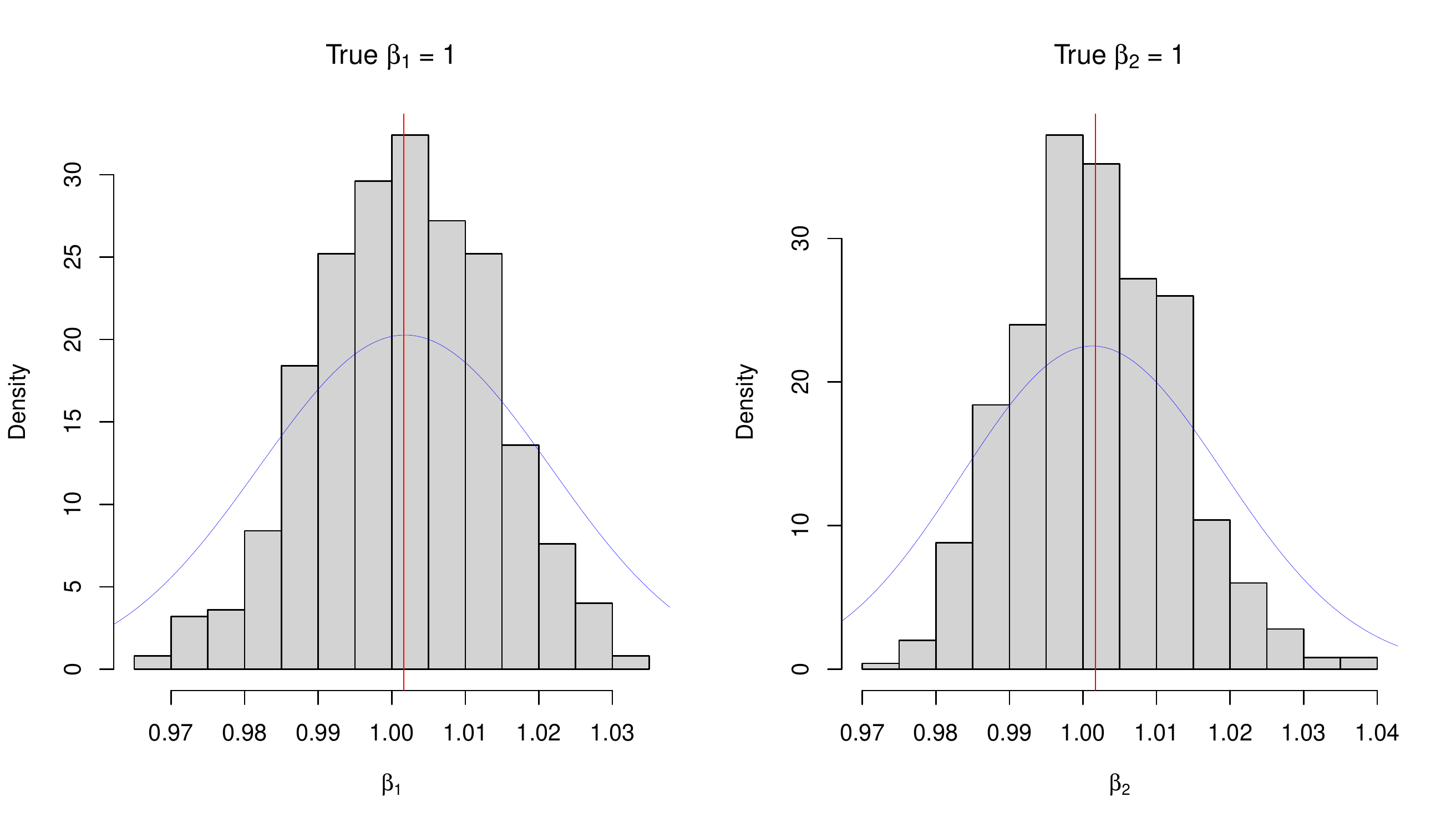}
	\caption{Histogram of regression parameters, $ \beta_1, \beta_2 $ for Case~$ 2 $ added by empirical density. The red line is the sample average. } \label{fig:case2}
\end{figure}

\begin{figure}[htbp!]
	\hspace{-10mm}%
	\includegraphics[width=1.2\textwidth]{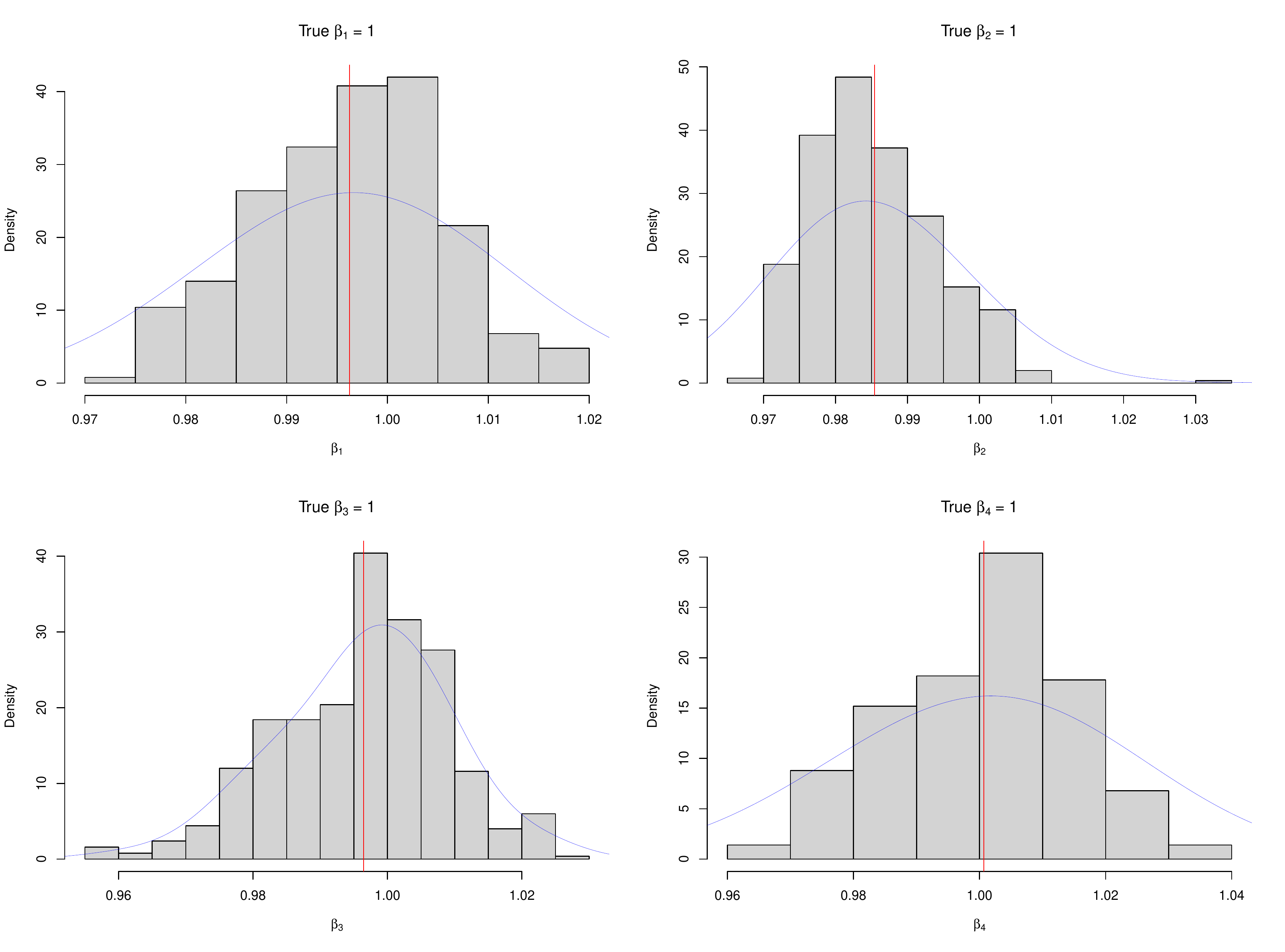}
	\caption{Histogram of regression parameters, $ \beta_1, \beta_2, \beta_3, \beta_4 $ for Case~$ 4 $ added by empirical density. The red line is the sample average.} \label{fig:case4}
\end{figure}

\begin{table}
	\caption{Case $5$: Estimated Parameters} \label{tab:case5_estimate}
	\centering
	\begin{tabular}{|c|c|c|c|c|}
		\hline
		& TRUE & HMC & L-BFGS-B & SQP \\ \hline
		$\alpha_1$ & 0.5 & 0.500 & 0.326 & 0.415 \\ \hline
		$\alpha_2$ & 0.5 & 0.500 & 0.834 & 0.419 \\ \hline
		$k_1$ & 0.2 & 0.191 & 0.481 & 0.955 \\ \hline
		$k_2$ & 0.2 & 0.195 & 0.187 & 0.030 \\ \hline
		$\lambda_1$ & 0.8 & 0.795 & 0.856 & 0.790 \\ \hline
		$\lambda_2$ & 0.8 & 0.814 & 0.172 & 0.276 \\ \hline
		$\beta_1$ & 1.0 & 1.003 & 0.108 & 1.977 \\ \hline
		$\beta_2$ & 1.0 & 1.003 & 1.528 & 0.968 \\ \hline
		$\gamma_{01}$ (intercept for $g=1$) & 1.0 & 1.010 & 0.326 & 0.687 \\ \hline
		$\gamma_{02}$ (intercept for $g=2$) & 1.0 & 0.995 & 1.520 & 1.568 \\ \hline
		$\gamma_1$ & 1.0 & 0.992 & 1.843 & 3.007 \\ \hline
		$\eta_1^2$ & 0.25 & 0.259 & 0.825 & 0.199 \\ \hline
		$\eta_1^2$ & 0.25 & 0.248 & 0.155 & 0.259 \\ \hline
		$\xi_1^2$ & 0.25 & 0.239 & 0.431 & 0.173 \\ \hline
		$\beta_{11}$ & 1.0 & 1.004 & 0.534 & 2.131 \\ \hline
		$\beta_{21}$ & 1.0 & 0.997 & 1.973 & 0.788 \\ \hline
		$\gamma_{11}$ & 1.0 & 0.948 & 0.891 & 0.904 \\ \hline
		$\beta_{12}$ & 1.0 & 1.011 & 1.225 & 0.253 \\ \hline
		$\beta_{22}$ & 1.0 & 0.995 & 1.558 & 0.208 \\ \hline
		$\gamma_{12}$ & 1.0 & 1.009 & 0.003 & 0.193 \\ \hline
		$\sigma^2 $ & 0.25 & 0.224 & 0.460 & 0.389 \\ \hline
	\end{tabular}
\end{table}

\begin{table}
		\caption{Case $6$: Estimated Parameters} \label{tab:case6_estimate}
	\centering
	\begin{tabular}{|c|c|c|c|c|}
		\hline
		& TRUE & HMC & L-BFGS-B & SQP \\ \hline
		$\alpha_1$ & 0.5 & 0.499 & 0.802 & 0.519 \\ \hline
		$\alpha_2$ & 0.5 & 0.498 & 0.662 & 0.619 \\ \hline
		$k_1$ & 0.2 & 0.211 & 0.132 & 0.324 \\ \hline
		$k_2$ & 0.2 & 0.199 & 0.371 & 0.068 \\ \hline
		$\lambda_1$ & 0.8 & 0.802 & 0.726 & 0.606 \\ \hline
		$\lambda_2$ & 0.8 & 0.809 & 0.437 & 0.444 \\ \hline
		$\beta_1$ & 1.0 & 0.998 & 1.309 & 2.288 \\ \hline
		$\beta_1$ & 1.0 & 0.998 & 1.532 & 0.515 \\ \hline
		$ \gamma_{01} $  (intercept for $g=1$) & 1.0 & 1.000 & 0.934 & 0.557 \\ \hline
		$ \gamma_{02} $  (intercept for $g=2$) & 1.0 & 1.001 & 1.482 & 1.561 \\ \hline
		$\gamma_1$ & 1.0 & 1.000 & 1.003 & 2.829 \\ \hline
		$\eta_1^2$ & 0.25 & 0.246 & 0.677 & 0.426 \\ \hline
		$\eta_2^2$ & 0.25 & 0.252 & 0.609 & 0.416 \\ \hline
		$\xi_1^2$ & 0.25 & 0.246 & 0.642 & 0.322 \\ \hline
		$\beta_{11}$ & 1.0 & 1.002 & 1.066 & 2.252 \\ \hline
		$\beta_{21}$ & 1.0 & 1.007 & 0.580 & 0.493 \\ \hline
		$\gamma_{11}$ & 1.0 & 0.946 & 0.956 & 0.920 \\ \hline
		$\beta_{12}$ & 1.0 & 1.002 & 1.097 & 0.352 \\ \hline
		$\beta_{22}$ & 1.0 & 1.011 & 0.315 & 1.290 \\ \hline
		$\gamma_{12}$ & 1.0 & 1.006 & 0.110 & 0.420 \\ \hline
		$\sigma^2 $ & 0.25 & 0.249 & 0.960 & 0.834 \\ \hline
	\end{tabular}
\end{table}

\begin{table}
	\caption{Case $7$: Estimated Parameters} \label{tab:case7_estimate}
	\hspace{-30mm}%
	\begin{tabular}{|c|c|c|c|c||c|c|c|c|c|}
		\hline
		& TRUE & HMC & L-BFGS-B & SQP &  & TRUE & HMC & L-BFGS-B & SQP \\ \hline
		$\alpha_1$ & 0.5 & 0.497 & 0.938 & 0.561 & $\eta_1^2$ & 0.25 & 0.255 & 0.913 & 0.474 \\ \hline
		$\alpha_2$ & 0.5 & 0.501 & 0.894 & 0.380 & $\eta_2^2$ & 0.25 & 0.243 & 0.967 & 0.601 \\ \hline
		$\alpha_3$ & 0.5 & 0.500 & 0.918 & 0.498 & $\eta_3^2$ & 0.25 & 0.238 & 0.714 & 0.078 \\ \hline
		$\alpha_4$ & 0.5 & 0.502 & 0.615 & 0.119 & $\eta_4^2$ & 0.25 & 0.244 & 0.462 & 0.099 \\ \hline
		$k_1$ & 0.2 & 0.202 & 0.342 & 0.508 & $\xi_1^2$ & 0.3 & 0.254 & 0.304 & 0.438 \\ \hline
		$k_2$ & 0.2 & 0.201 & 0.166 & 0.523 & $\beta_{11}$ & 1.0 & 0.993 & 1.203 & 0.471 \\ \hline
		$k_3$ & 0.2 & 0.200 & 0.906 & 0.053 & $\beta_{21}$ & 1.0 & 1.001 & 0.412 & 3.301 \\ \hline
		$k_4$ & 0.2 & 0.194 & 0.628 & 0.507 & $\beta_{31}$ & 1.0 & 0.995 & 1.963 & 0.440 \\ \hline
		$\lambda_1$ & 0.8 & 0.798 & 0.844 & 0.364 & $\beta_{41}$ & 1.0 & 1.009 & 0.791 & 0.159 \\ \hline
		$\lambda_2$ & 0.8 & 0.808 & 0.201 & 0.485 & $\gamma_{11}$ & 1.0 & 0.958 & 0.985 & 0.929 \\ \hline
		$\lambda_3$ & 0.8 & 0.788 & 0.106 & 0.463 & $\beta_{12}$ & 1.0 & 1.010 & 1.396 & 0.935 \\ \hline
		$\lambda_4$ & 0.8 & 0.807 & 0.759 & 0.055 & $\beta_{22}$ & 1.0 & 1.016 & 0.230 & 1.751 \\ \hline
		$\beta_1$ & 1.0 & 1.005 & 0.524 & 0.219 & $\beta_{32}$ & 1.0 & 1.005 & 0.995 & 0.620 \\ \hline
		$\beta_2$ & 1.0 & 1.002 & 1.038 & 2.309 & $\beta_{42}$ & 1.0 & 1.010 & 0.613 & 0.071 \\ \hline
		$\beta_3$ & 1.0 & 1.000 & 1.790 & 0.412 & $\gamma_{21}$ & 1.0 & 0.994 & 1.027 & 1.978 \\ \hline
		$\beta_4$ & 1.0 & 1.004 & 1.928 & 0.212 & $\sigma^2 $ & 0.25 & 0.246 & 0.033 & 0.495 \\ \hline
		$ \gamma_{01} $  (intercept for $g=1$) & 1.0 & 1.003 & 0.402 & 0.983 &  &  &  &  &  \\ \hline
		$ \gamma_{02} $  (intercept for $g=2$) & 1.0 & 1.011 & 1.711 & 0.885 &  &  &  &  &  \\ \hline
		$\gamma_1$ & 1.0 & 1.004 & 0.553 & 0.258 &  &  &  &  &  \\ \hline
	\end{tabular}
\end{table}

\begin{table}
	\caption{Case $8$: Estimated Parameters} \label{tab:case8_estimate}
	\hspace{-30mm}%
	\begin{tabular}{|c|c|c|c|c||c|c|c|c|c|}
		\hline
		& TRUE & HMC & L-BFGS-B & SQP &  & TRUE & HMC & L-BFGS-B & SQP \\ \hline
		$\alpha_1$ & 0.5 & 0.499 & 0.181 & 0.810 & $\eta_1^2$ & 0.25 & 0.237 & 0.146 & 0.467 \\ \hline
		$\alpha_2$ & 0.5 & 0.502 & 0.158 & 0.646 & $\eta_2^2$ & 0.25 & 0.244 & 0.166 & 0.233 \\ \hline
		$\alpha_3$ & 0.5 & 0.500 & 0.393 & 0.921 & $\eta_3^2$ & 0.25 & 0.241 & 0.832 & 0.082 \\ \hline
		$\alpha_4$ & 0.5 & 0.502 & 0.948 & 0.490 & $\eta_4^2$ & 0.25 & 0.243 & 0.037 & 0.452 \\ \hline
		$k_1$ & 0.2 & 0.207 & 0.453 & 0.541 & $\xi_1^2$ & 0.25 & 0.250 & 0.493 & 0.356 \\ \hline
		$k_2$ & 0.2 & 0.191 & 0.878 & 0.864 & $\beta_{11}$ & 1.0 & 0.988 & 0.467 & 1.376 \\ \hline
		$k_3$ & 0.2 & 0.199 & 0.693 & 0.298 & $\beta_{21}$ & 1.0 & 0.993 & 1.110 & 0.601 \\ \hline
		$k_4$ & 0.2 & 0.205 & 0.182 & 0.124 & $\beta_{31}$ & 1.0 & 0.987 & 1.691 & 1.536 \\ \hline
		$\lambda_1$ & 0.8 & 0.798 & 0.533 & 0.603 & $\beta_{41}$ & 1.0 & 0.999 & 0.205 & 0.804 \\ \hline
		$\lambda_2$ & 0.8 & 0.793 & 0.764 & 0.109 & $\gamma_{11}$ & 1.0 & 1.003 & 0.962 & 1.001 \\ \hline
		$\lambda_3$ & 0.8 & 0.803 & 0.793 & 0.315 & $\beta_{12}$ & 1.0 & 0.993 & 0.437 & 1.935 \\ \hline
		$\lambda_4$ & 0.8 & 0.789 & 0.923 & 0.720 & $\beta_{22}$ & 1.0 & 1.005 & 0.763 & 0.869 \\ \hline
		$\beta_1$ & 1.0 & 0.999 & 1.359 & 1.787 & $\beta_{32}$ & 1.0 & 1.006 & 0.625 & 1.088 \\ \hline
		$\beta_2$ & 1.0 & 1.003 & 0.362 & 1.198 & $\beta_{42}$ & 1.0 & 1.004 & 0.031 & 0.991 \\ \hline
		$\beta_3$ & 1.0 & 0.997 & 0.156 & 1.178 & $\gamma_{21}$ & 1.0 & 0.996 & 1.273 & 0.362 \\ \hline
		$\beta_4$ & 1.0 & 1.007 & 1.605 & 1.436 & $\sigma^2 $ & 0.3 & 0.253 & 0.451 & 0.548 \\ \hline
		$ \gamma_{01} $  (intercept for $g=1$) & 1.0 & 0.984 & 1.043 & 0.499 &  &  &  &  &  \\ \hline
		$ \gamma_{02} $  (intercept for $g=2$) & 1.0 & 1.014 & 0.910 & 1.195 &  &  &  &  &  \\ \hline
		$\gamma_1$ & 1.0 & 0.999 & 0.107 & 0.452 &  &  &  &  &  \\ \hline
	\end{tabular}
\end{table}

\begin{figure}[htbp!]
	\hspace{-10mm}%
	\includegraphics[width=1.2\textwidth]{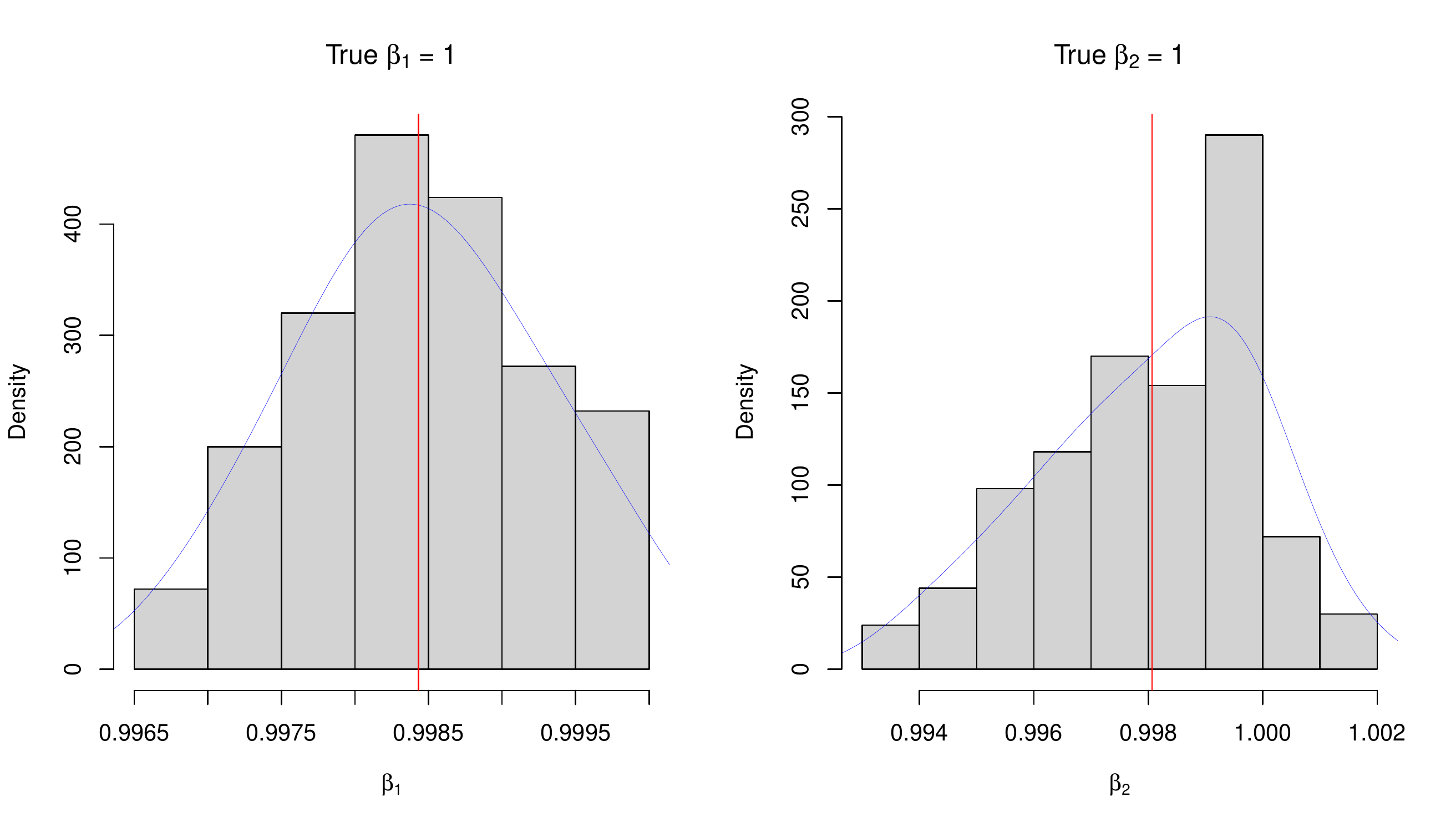}
	\caption{Histogram of fixed effects, $ \beta_1, \beta_2 $ for Case~$ 6 $ added by empirical density. The red line is the sample average.} \label{fig:case6_fe}
\end{figure}

\begin{figure}[htbp!]
	\hspace{-10mm}%
	\includegraphics[width=1.2\textwidth]{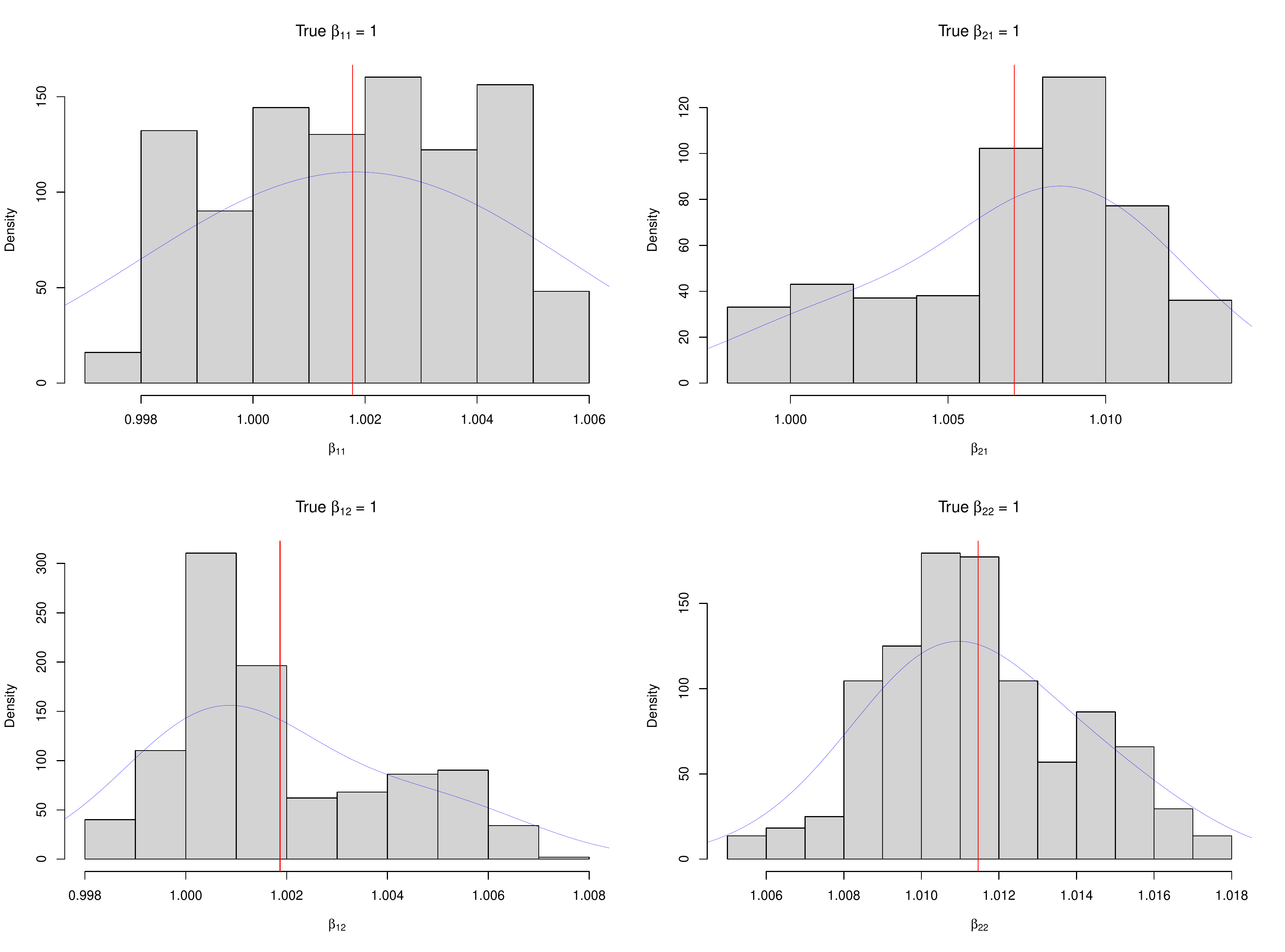}
	\caption{Histogram of random effects, $ \beta_{1, 1}, \beta_{2, 1}, \beta_{1, 2}, \beta_{2, 2} $ for Case~$ 6 $ added by empirical density. The red line is the sample average.} \label{fig:case6_rf}
\end{figure}

\begin{figure}[htbp!]
	\hspace{-10mm}%
	\includegraphics[width=1.2\textwidth]{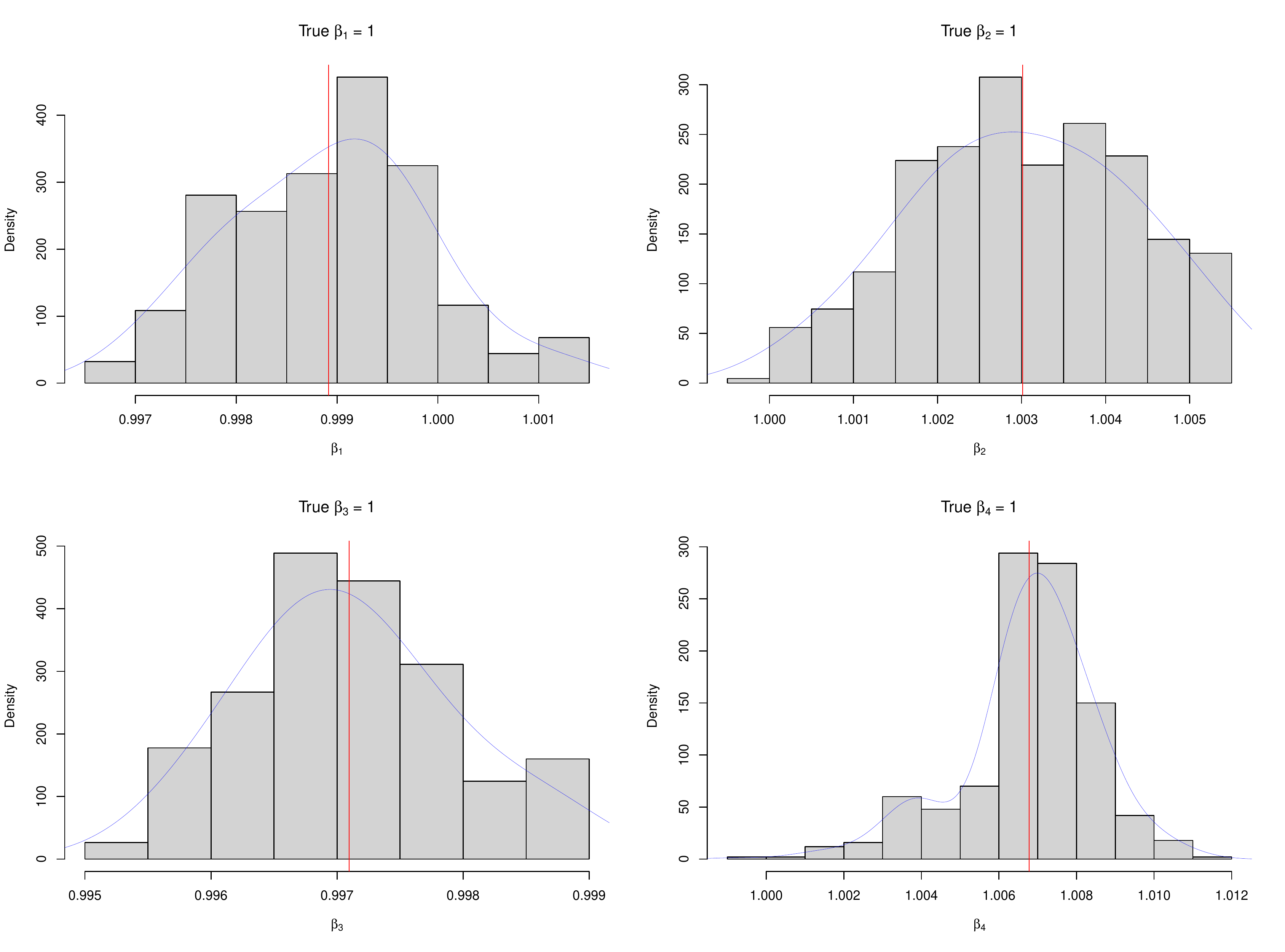}
	\caption{Histogram of regression parameters, $ \beta_1, \beta_2, \beta_3, \beta_4 $ for Case~$ 8 $ added by empirical density. The red line is the sample average.} \label{fig:case8}
\end{figure}

\begin{table}[htbp!]
	\caption {Estimates from HMC for the real application 2}  \label{tab:g3}
	\begin{center}
		\begin{tabular}{|cc|cc|}\hline
			Parameters  & HMC & Parameters  & HMC   \\ \hline \hline
			$\alpha_1$     & 0.75  &  $ \beta_{1, 1} $ & 0.04   \\ \hline
			$\alpha_2$     & 0.72     &  $ \beta_{2, 1} $ & 0.25   \\ \hline
			$\alpha_3$       & 0.71     & $ \beta_{3, 1} $ & 0.03   \\ \hline
			$\alpha_4$       & 0.70     & $ \beta_{4, 1} $ & 0.25   \\ \hline 	
			$k_1$     &  0.50     &  $ \gamma_{1, 1} $ & 0.23   \\ \hline
			$k_2$     & 0.36   &  $ \gamma_{2, 1} $ & 0.02   \\ \hline
			$k_3$     &  0.33   &  $ \gamma_{3, 1} $ & 0.04   \\ \hline
			$k_4$     &  0.45   &  $ \beta_{2, 1} $ & 0.05   \\ \hline 		
			$\lambda_1$  & 0.38         & $ \beta_{2, 2} $ & 0.24  \\ \hline
			$\lambda_2$  &  0.48        &  $ \beta_{3, 2} $ & 0.04   \\ \hline
			$\lambda_3$  &  0.25       &  $ \beta_{4, 2} $ & 0.24   \\ \hline
			$\lambda_4$  &  0.46       &  $ \gamma_{1, 2} $ & 0.26   \\ \hline		
			$\beta_1$    &  0.27   & $ \gamma_{2, 2} $ & 0.01   \\ \hline 		
			$\beta_2$       &  0.24   & $ \gamma_{3, 2} $ & 0.03   \\ \hline 		
			$\beta_3$      &  0.26   & $ \beta_{1, 3} $ & 0.03   \\ \hline
			$\beta_4$      &  0.25   & $ \beta_{2, 3} $ & 0.25   \\ \hline	
			$\gamma_1$       & 0.25    &  $ \beta_{3, 3} $ & 0.03    \\ \hline
			$\gamma_2$       & 0.23    &  $ \beta_{4, 3} $ & 0.24    \\ \hline
			$\gamma_3$       & 0.25    &  $ \gamma_{1, 3} $ & 0.24    \\ \hline 		
			$\eta_{1}^2$    & 0.25  & $ \gamma_{2, 3} $ & 0.02   \\ \hline 	
			$\eta_{2}^2$    & 0.26  & $ \gamma_{3, 3} $ & 0.02   \\ \hline 	
			$\eta_{3}^2$    & 0.24  & $ \beta_{1, 4} $ & 0.02   \\ \hline
			$\eta_{4}^2$    & 0.19  & $ \beta_{2, 4} $ & 0.24   \\ \hline 	
			$\xi_{1}^2$    &  0.24  &  $\beta_{3, 4}$   &  0.03      \\ \hline
			$\xi_{2}^2$    &  0.23  &  $\beta_{4, 4}$   &  0.25      \\ \hline
			$\xi_{3}^2$    &  0.23  &  $\gamma_{1, 4}$   &  0.19      \\ \hline
			$\sigma^2$    &  1.59  &  $ \gamma_{2, 4}  $   &  0.02      \\ \hline
			&   &  $\gamma_{3, 4} $   &  0.03      \\ \hline	
		\end{tabular}
	\end{center}
\end{table}

\end{document}